\documentclass[trackchanges,twocolumn]{aastex701}

\DeclareRobustCommand{\ion}[2]{\mbox{#1\,\textsc{#2}}}
\usepackage[flushleft]{threeparttable}
\usepackage{placeins}
\usepackage{multirow}
\usepackage{amsmath}
\usepackage{ulem}
\usepackage{cancel}

\begin{document}


\title{Two Peas in a Pod: The First Confirmed Dual Active Galactic Nucleus within a Green Pea Galaxy System}

\author[0000-0002-1444-2016]{Konstantinos~Kouroumpatzakis}
\email{konstantinos.kouroumpatzakis@asu.cas.cz}
\affiliation{Astronomical Institute, Academy of Sciences, Bo\v{c}n\'{i} II 1401, CZ-14131 Prague, Czech Republic}

\author[0000-0001-9379-4716]{Peter~G.~Boorman}
\email{boorman@mpe.mpg.de}
\affiliation{Max Planck Institute for Extraterrestrial Physics, Giessenbachstrasse, 85741 Garching, Germany}
\affiliation{Cahill Center for Astrophysics, California Institute of Technology, 1216 East California Boulevard, Pasadena, CA 91125, USA}

\author[0000-0003-2931-0742]{Ji\v{r}\'{i}~Svoboda}
\email{jiri.svoboda@asu.cas.cz}
\affiliation{Astronomical Institute, Academy of Sciences, Bo\v{c}n\'{i} II 1401, CZ-14131 Prague, Czech Republic}

\author[0000-0001-8640-8522]{Ryan Pfeifle}
\email{rpfeifle@gmu.edu}
\affiliation{U.S. Naval Observatory, 3450 Massachusetts Avenue NW, Washington, DC 20392, USA}

\author[0000-0002-9807-4520]{Abhijeet Borkar}
\email{abhijeet.borkar@asu.cas.cz}
\affiliation{Astronomical Institute, Academy of Sciences, Bo\v{c}n\'{i} II 1401, CZ-14131 Prague, Czech Republic}

\author[0000-0003-0976-8932]{Maitrayee Gupta}
\email{maitrayee.gupta@asu.cas.cz}
\affiliation{Astronomical Institute, Academy of Sciences, Bo\v{c}n\'{i} II 1401, CZ-14131 Prague, Czech Republic}

\author[0000-0003-2686-9241]{Daniel Stern}
\email{daniel.k.stern@jpl.nasa.gov}
\affiliation{Cahill Center for Astronomy and Astrophysics, California Institute of Technology, 1216 E California Blvd, Pasadena, CA 91125, USA}
\affiliation{Jet Propulsion Laboratory, California Institute of Technology, Pasadena, CA 91109, USA}

\author[0000-0001-8952-676X]{Andreas Zezas}
\email{azezas@physics.uoc.gr}
\affiliation{Physics Department - University of Crete, GR 71003, Heraklion, Greece}
\affiliation{Institute of Theoretical and Computational Physics - University of Crete, GR 71003, Heraklion, Greece}
\affiliation{Institute of Astrophysics, Foundation for Research and Technology-Hellas, GR 71110 Heraklion, Greece}


\begin{abstract}
The growth of galaxies in the early Universe is thought to be dominated by compact, intensely star-forming systems, but the growth of supermassive black holes (SMBHs) in these environments remains poorly constrained.
Green Pea galaxies are nearby analogs of rapidly assembling galaxies, owing to their compact morphologies, intense star formation, low metallicities, and extreme ionization conditions.
Although interactions may trigger rapid SMBH growth, direct evidence for simultaneous accretion onto multiple SMBHs in such systems is lacking.
Here we report the first confirmed dual active galactic nucleus (AGN) in a Green Pea galaxy, SDSS~J162209.41+352107.5.
\textit{Chandra} imaging resolves two luminous hard X-ray sources separated by 8.4~kpc in projection, demonstrating simultaneous accretion onto two SMBHs.
Keck spectroscopy confirms broad Balmer emission and AGN-like narrow-line ratios in both nuclei.
High-ionization emission lines, including [\ion{Fe}{VII}] and [\ion{Ne}{V}], are also detected in the optical spectra.
The unresolved UV-to-optical continuum exhibits a V-shaped spectral energy distribution, with a UV slope within the range commonly reported for Little Red Dots and a positive but relatively shallow optical slope, adding to the phenomenological similarities between this system and compact broad-line AGN populations identified at high redshift.
Unlike most known dual AGN, typically found in massive mergers, J162209.41+352107.5 is a compact system with host-galaxy properties resembling those of galaxies in the early stages of assembly.
This discovery shows that multiple SMBHs can grow efficiently in such environments and establishes Green Pea galaxies as nearby laboratories for studying the interplay between galaxy interactions, star formation, and black-hole growth.
\end{abstract}



\section{Introduction}
\label{sec:introduction}

The growth of galaxies in the early Universe is thought to be dominated by compact, intensely star-forming systems, yet the corresponding growth of supermassive black holes (SMBHs) in these environments remains poorly constrained \citep[e.g.,][]{2023ApJS..265....5H,2024ApJ...964...39G}.
Nearby analogs provide an opportunity to investigate the physical processes that regulate star formation and SMBH growth at higher spatial resolution than is generally possible for comparable systems at high redshift.

Green Pea (GP) galaxies are among the best-studied examples of such local systems.
Originally identified in the Sloan Digital Sky Survey (SDSS) through their strong [\ion{O}{III}]~$\lambda5007$ emission \citep{2009MNRAS.399.1191C}, they are compact, low-metallicity galaxies with extreme ionization conditions and high specific star-formation rates \citep[${\rm sSFR \equiv SFR}/M_{\star} $; e.g.,][]{2010ApJ...715L.128A,2011ApJ...728..161I,2017ApJ...838....4Y,2021ApJ...914....2K,2024A&A...688A.159K}.
A subset of GPs host active galactic nuclei \citep[AGN;][]{2009MNRAS.399.1191C,2019ApJ...880..144S,2023MNRAS.524.2224L,2025ApJ...980L..34L,2025ApJ...988..157B}, demonstrating that accretion activity can occur in compact galaxies undergoing intense star formation.

Dual AGN, in which two SMBHs accrete simultaneously within an interacting or merging galaxy system, trace an important phase of galaxy evolution and SMBH growth \citep[e.g.,][]{2019NewAR..8601525D,2023MNRAS.522.1895C,2025ApJS..281...25P}.
They provide a means of studying SMBH growth during galaxy interactions and the stages preceding the formation of close SMBH binaries.
Despite extensive multiwavelength searches, confirmed dual AGN remain relatively rare and are predominantly found in massive, gas-rich mergers at $z\lesssim1$ \citep[e.g.,][]{2011ApJ...735...48S,2017ApJ...848..126S,2023MNRAS.519.5149D,2025ApJS..281...25P}.
Several GPs have been reported to show double-peaked [\ion{O}{III}] emission suggestive of dual AGN activity \citep{2023MNRAS.524.2224L}.
However, such profiles can also arise from outflows or complex gas kinematics and are not, by themselves, definitive evidence for two accreting nuclei \citep[e.g.,][]{2011ApJ...733..103F,2012ApJ...745...67F,2010ApJ...708..419C,2015ApJ...811...14M}.
To our knowledge, no dual AGN had previously been confirmed in a compact, metal-poor GP system \citep[e.g.,][]{2025ApJS..281...25P}.

The discovery of compact AGN populations with the \textit{James Webb Space Telescope} (\textit{JWST}), including the so-called ``Little Red Dots'' (LRDs) and ``Little Blue Dots'' (LBDs), has renewed interest in the relationship between early SMBH growth and compact star-forming environments \citep[e.g.,][]{2023ApJS..265....5H,2024ApJ...976...96P,2024ApJ...963..129M,2025ApJ...986..126K,2026arXiv260322277S}.
The physical nature of these populations remains under active investigation, and their relationship to local compact galaxies is not yet fully established.
Nevertheless, some nearby GPs exhibit spectroscopic and interstellar-medium properties similar to those inferred for compact star-forming galaxies in the early Universe \citep[e.g.,][]{2022A&A...665L...4S,2023ApJ...942L..14R,2025ApJ...980L..34L}.
These similarities motivate the use of GPs as local laboratories for investigating the interplay between intense star formation, nuclear activity, and the conditions associated with rapid galaxy assembly, without requiring a one-to-one evolutionary correspondence between the local and high-redshift populations.

In this work, we investigate SDSS~J162209.41+352107.5 (hereafter J1622+3521), a GP galaxy previously classified as a narrow-line Seyfert~1 (NLS1) and selected for follow-up because of its disturbed optical morphology and potential nuclear activity.
We combine \textit{Chandra} X-ray observations with spatially resolved Keck/DEIMOS spectroscopy to investigate its nuclear structure and the nature of its two optical components.

The paper is organized as follows. Section~\ref{sec:data_methods} describes the observations, data reduction, and spectral analysis; Section~\ref{sec:results} presents the X-ray and optical results; Section~\ref{sec:Discussion} discusses their implications in the context of dual AGN and compact high-redshift AGN populations; and Section~\ref{sec:conclusion} summarizes our findings.
Throughout this study, we adopt a flat $\Lambda$CDM cosmology \citep{2020A&A...641A...6P} with $\Omega_m = 0.315$ and $h = 0.674$, and uncertainties correspond to the 68\% confidence level unless otherwise stated.

\section{Data, Methods, and Results}
\label{sec:data_methods}

We perform an X-ray spectral and broadband analysis of J1622+3521 using 42.7 ks of newly obtained data from three \textit{Chandra} observations (PI: K.~Kouroumpatzakis; 
Observation IDs: 28142, 30637, and 30638).
The X-ray analysis is complemented by optical long-slit spectroscopic observations acquired with Keck/DEIMOS.
We also use archival DESI Legacy Imaging Survey \citep[DeCaLs;][]{2019AJ....157..168D} and SDSS data.

\begin{figure*}[ht!]
    \centering
    \includegraphics[width=0.7\linewidth]{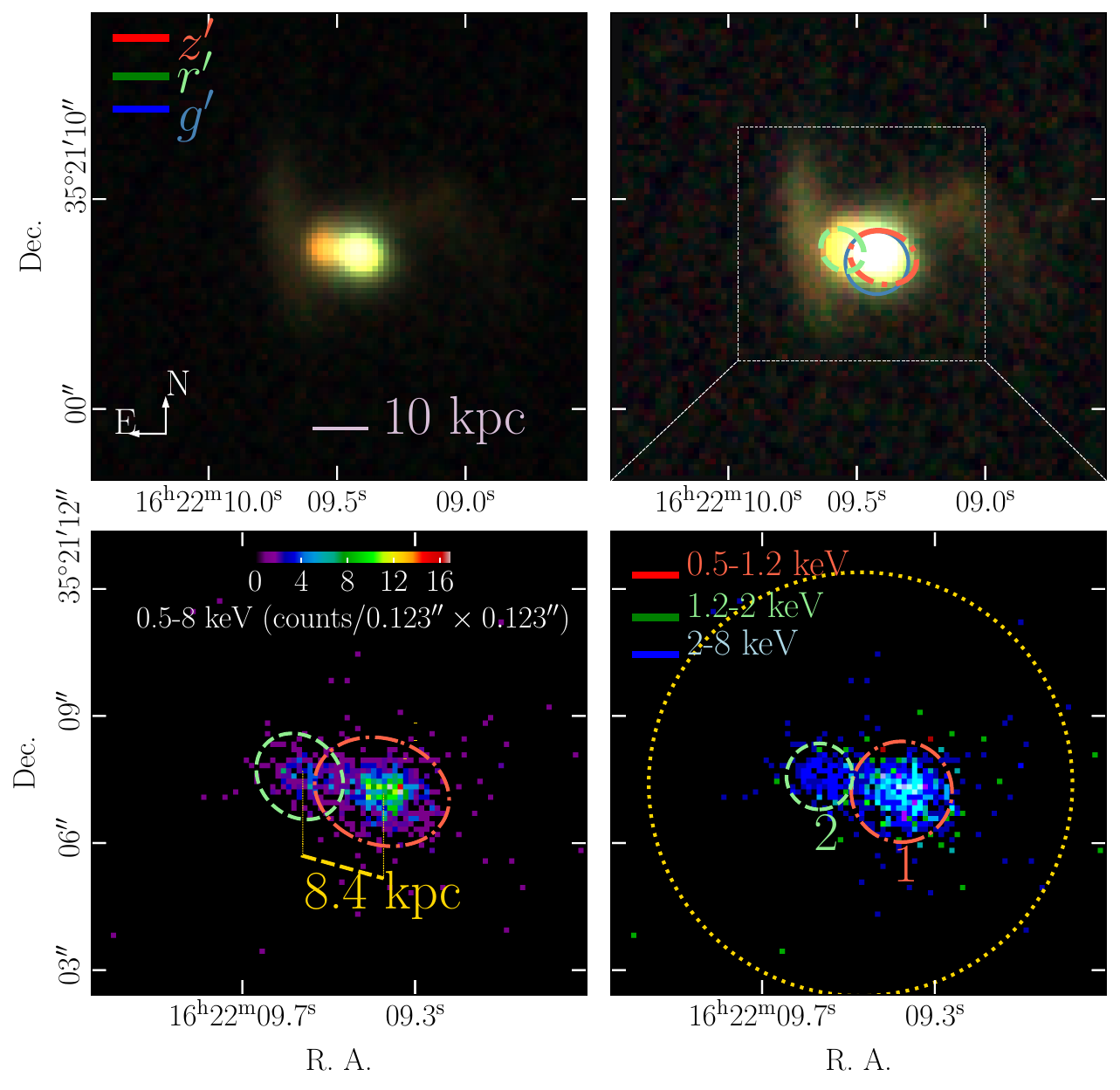}
    \caption{
    Optical and X-ray views of J1622+3521.
    (\textit{Upper left}) DECaLS optical composite image combining the $z$, $r$, and $g$ bands mapped to RGB colors.
    The pink bar indicates a projected physical scale of 10~kpc.
    (\textit{Upper right}) The same DECaLS image is shown with enhanced contrast to highlight faint filamentary structures and tidal features associated with the ongoing interaction.
    The colored ellipses mark the source-detection regions of the two X-ray sources identified in the \textit{Chandra} image (lower left), while the blue circle outlines the SDSS spectroscopic fiber.
    (\textit{Lower left}) \textit{Chandra} 0.5--8.0~keV image of the central region, showing two spatially resolved X-ray sources.
    The color scale represents the X-ray intensity.
    The yellow line marks the projected separation of 8.4~kpc between the X-ray source centroids.
    (\textit{Lower right}) Three-color \textit{Chandra} image with the soft (0.5--1.2~keV), medium (1.2--2.0~keV), and hard (2.0--8.0~keV) bands mapped to RGB colors.
    The red and green circles indicate the X-ray spectral extraction regions for Sources~1 and 2, respectively, while the yellow dashed circle marks the inner boundary of the background region.
    }
    \label{fig:Chandra_PS1_images}
\end{figure*}

\subsection{Optical Data}
\label{sec:Optical_data}

Deep optical imaging from the DECaLS survey reveals a disturbed morphology characteristic of an advanced merger, with two distinct optical nuclei separated by diffuse stellar emission (Figure~\ref{fig:Chandra_PS1_images}).
The western component appears significantly bluer than the eastern one.
In addition, faint asymmetric and filamentary structures extend predominantly along the east--west direction, consistent with tidal debris produced by an ongoing galaxy interaction.
The positions of the two optical nuclei are consistent with those of the X-ray sources detected by \textit{Chandra}.

Follow-up optical spectroscopy was obtained with the DEIMOS spectrograph on the Keck~II telescope on UT 2025 February 22. 
The observations were carried out in long-slit mode using the 600ZD grating and the GG400 order-blocking filter, providing a wavelength coverage of $\simeq 4750$--$10050$~\AA\ with a central wavelength of $\sim7400$~\AA. 
The slit mask (``Long1.0B'') corresponds to a slit width of $1''$, yielding a spectral resolution of $R \simeq 1370$ and a dispersion of $\simeq 0.65$~\AA\,pix$^{-1}$.
The observations were conducted at an airmass of $\simeq 1.2$ under good conditions, with a guide-star FWHM of $\simeq 0.88''$.
The total integration time was 900~s, and the slit was oriented to simultaneously cover both optical counterparts associated with the X-ray emission.

\begin{figure*}[ht!]
\centering
\includegraphics[width=1\linewidth]{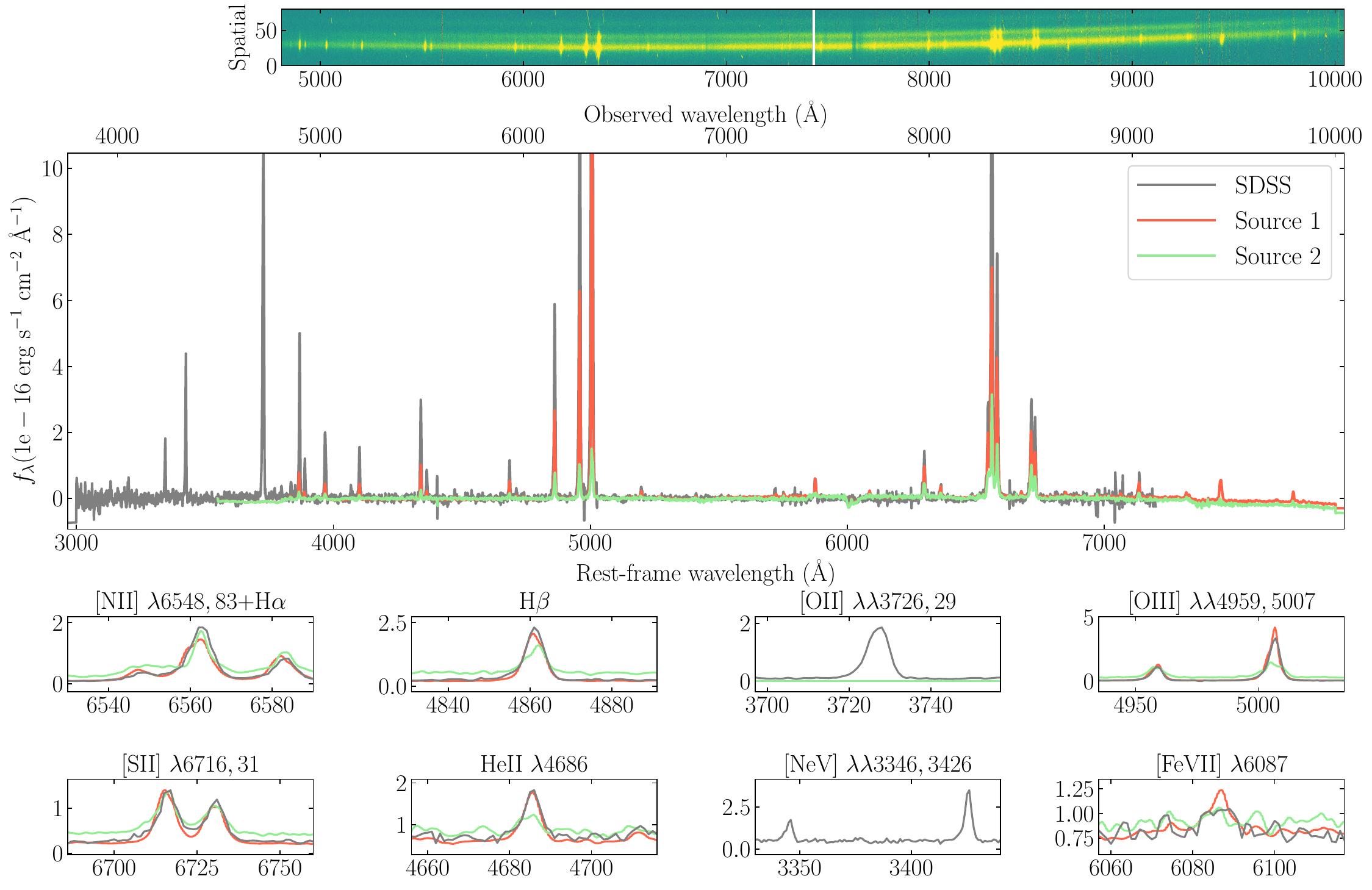}
\caption{
Optical spectra of the X-ray-bright Source~1 (red) and Source~2 (green), obtained with Keck/DEIMOS, together with the SDSS fiber spectrum (gray).
The upper panel shows the observed-frame two-dimensional spectrum, the middle panel the extracted rest-frame spectra, and the lower panels representative emission-line regions.
}
\label{fig:Keck_speck}
\end{figure*}

The data were reduced using the \texttt{PypeIt} pipeline \citep{2020JOSS....5.2308P}, which performs bias subtraction, flat-fielding, wavelength calibration, sky subtraction, and optimal extraction of one-dimensional spectra. 
Flux calibration was performed using observations of the spectrophotometric standard star HZ44 obtained during the same night.
The resulting two-dimensional spectrum (Figure~\ref{fig:Keck_speck}, top panel) clearly shows two spatially separated emission components along the slit, corresponding to the two optical nuclei. 
Both components exhibit strong emission lines across the full spectral range. 
Although the data were rectified using the \texttt{PypeIt} pipeline, small residual geometric distortions remain visible in the two-dimensional spectra, particularly toward the detector edges. 
Visual inspection confirms that the tracing algorithm successfully follows the continuum and emission-line structure across the full wavelength range for both components, and therefore these residual distortions do not affect the wavelength calibration or the extracted one-dimensional spectra used for the emission-line analysis.

Given the projected separation of only $\sim1.9''$, some low-level spill-over between the extraction apertures is expected, particularly in the brightest emission lines.
However, the two-dimensional spectrum clearly reveals distinct emission-line structures associated with each nucleus.
The corresponding emission-line features in the two nuclei are centered at consistent observed wavelengths, indicating that the two components share a common redshift.
In features such as [\ion{O}{III}]~$\lambda5007$ and \ion{H}{$\alpha$}, the spatial profiles become narrower between the two continuum traces and broaden again at the position of the second nucleus, demonstrating that the line emission is spatially dominated by the respective sources rather than by cross-contamination.
The spatial structure of the emission lines therefore supports the interpretation that the two nuclei are distinct sources, while their measured redshifts are consistent with a common systemic velocity.

The one-dimensional spectra of the two components are shown in Figure~\ref{fig:Keck_speck} (middle panel) and are consistent with those observed in the SDSS fiber spectrum.
Both spectra exhibit strong emission lines, including the Balmer series, prominent [\ion{O}{III}]~$\lambda\lambda4959,5007$, [\ion{N}{II}]~$\lambda\lambda6548,6583$, and [\ion{S}{II}]~$\lambda\lambda6716,6731$ transitions.
Zoomed-in views of key diagnostic regions (Figure~\ref{fig:Keck_speck}, lower panels) show that the emission-line profiles are well resolved.
In addition, high-ionization coronal lines such as [\ion{Ne}{V}] $\lambda\lambda3346,3426$ and [\ion{Fe}{VII}] $\lambda6087$ are detected (Appendix \ref{app:high_ionization}), supporting the presence of an active nucleus.
Overall, the spectroscopic data confirm that both optical components are at consistent redshift and exhibit AGN-like emission-line properties, supporting the interpretation of a dual AGN system embedded in a merging host galaxy.

\subsection{Optical Spectral Analysis}
\label{sec:Optical_Spectral_Analysis}

To further constrain the sources' properties, we analyzed the continuum-subtracted optical spectra of J1622+3521 using Keck/DEIMOS spectra of Source~1 and Source~2, which have higher resolution than the SDSS spectrum.
The underlying continuum was first modeled and subtracted using \texttt{FADO} \citep{2017A&A...603A..63G}, which self-consistently accounts for both stellar and nebular continuum emission.
The resulting residual spectra were subsequently used for the emission-line analysis.

The DEIMOS spectra were transformed to an initial rest frame using the SDSS redshift of $z=0.2665$ from the MPA--JHU catalog \citep{2003MNRAS.346.1055K,10.1111/j.1365-2966.2004.07881.x,Tremonti_2004}. 
The systemic velocities of the individual nuclei were then obtained from the \texttt{FADO} modeling.
The best-fit \texttt{FADO} models yield redshifts of
$z_{\rm S1}=0.2667 \pm 0.0001$ and
$z_{\rm S2}=0.2667 \pm 0.0002$
for Source~1 and Source~2, respectively.
These values correspond to velocity offsets of
$\sim61 \pm 42$~km~s$^{-1}$ and
$\sim68 \pm 61$~km~s$^{-1}$ relative to the systemic SDSS redshift.
The implied velocity difference between the two nuclei is only $\sim7$~km~s$^{-1}$ and is not significant given the uncertainties.
Within the uncertainties, the two nuclei are therefore consistent with sharing a common systemic velocity, although modest internal kinematic differences associated with the ongoing interaction may still be present.
At this redshift, the corresponding luminosity and angular diameter distances are $1400$ and $873$~Mpc, respectively.
The astrometric properties of the two nuclei are summarized in Table~\ref{tab:dualAGN_geometry}. 

\begin{table}[t]
\centering
\caption{
Astrometric properties of the two nuclei in the source~J1622+3521.
Coordinates correspond to the centroids of the \textit{Chandra} X-ray detections obtained with {\tt wavdetect} (Section \ref{sec:X_ray_source_detection}), while redshifts are derived from the \texttt{FADO} spectral modeling of the DEIMOS spectra.
}
\label{tab:dualAGN_geometry}
\renewcommand{\arraystretch}{1.2}
\setlength{\tabcolsep}{4pt}
\begin{tabular}{lccc}
\hline
Source & R.A. & Decl. & Redshift \\
\hline
Source~1
& 16:22:09.40
& +35:21:07.2
& $0.2667 \pm 0.0001$ \\

Source~2
& 16:22:09.56
& +35:21:07.6
& $0.2667 \pm 0.0002$ \\

\end{tabular}
\end{table}

We modeled the continuum-subtracted spectra using the \texttt{Sherpa} fitting package within \texttt{CIAO} \citep{2001SPIE.4477...76F, SciPyProceedings_51}.
To account for uncertainties beyond the statistical errors, we adopted a total uncertainty of $\sigma_{\rm tot} = \sqrt{\sigma_{\rm stat}^{2} + (0.05|F|)^{2}}$, where $F$ is the measured flux density.
We fitted the \ion{H}{$\alpha$}, \ion{H}{$\beta$}, [\ion{O}{III}], and \ion{He}{II} emission-line regions separately, allowing for a local constant continuum in each region.
For each region, we compared three models (M1, M2, and M3) to assess whether additional kinematic components were required.
Model comparisons were based on the $\chi^2$ statistic, the degrees of freedom (DoF), and the approximate Bayesian information criterion (BIC), with lower BIC values indicating a preferred model.

The \ion{H}{$\alpha$} region ($6480$--$6660$~\AA) was fitted with Gaussian components representing the [\ion{N}{II}] $\lambda\lambda6548,6583$ doublet and \ion{H}{$\alpha$}.
We fixed the [\ion{N}{II}] line-flux ratio to its theoretical value of 2.96 and initially tied the widths of the narrow components, including that of \ion{H}{$\alpha$}.
Model M1 contains a single narrow component for the emission lines, M2 allows two narrow components, and M3 consists of a narrow component plus a broad \ion{H}{$\alpha$} component.
In M3, the narrow components describe the narrow-line emission, while the additional broad Gaussian accounts for the extended wings of \ion{H}{$\alpha$}.
The broad-component model is strongly favoured over the narrow-only alternatives for both sources.
For Source~1, M1 gives $\chi^2=2824.55$ for 714 DoF (BIC = 2857.44), M2 gives $\chi^2=1235.33$ for 710 DoF (BIC = 1294.53), and M3 gives $\chi^2=720.44$ for 707 DoF (BIC = 799.38).
For Source~2, the corresponding BIC values are 3167.1, 1920.0, and 1335.3 for M1, M2, and M3, respectively.
The broad \ion{H}{$\alpha$} components have FWHM values of approximately $1900$ and $1700$~km~s$^{-1}$ for Sources~1 and 2, respectively.
Their widths and the substantial improvement in fit quality support the presence of broad-line emission, consistent with a broad-line region in a narrow-line Seyfert~1 (NLS1) system.

The \ion{H}{$\beta$} region ($4830$--$4890$~\AA) was modelled using the same three configurations: M1 with one narrow Gaussian, M2 with two narrow Gaussians, and M3 with one narrow and one broad Gaussian.
For Source~1, M1 gives $\chi^2=1622.38$ for 475 DoF (BIC = 1676.69), M2 gives $\chi^2=982.77$ for 472 DoF (BIC = 1077.82), and M3 gives $\chi^2=588.12$ for 472 DoF (BIC = 683.17).
For Source~2, M1 gives $\chi^2=966.82$ for 475 DoF (BIC = 1021.13), M2 gives $\chi^2=718.86$ for 472 DoF (BIC = 813.91), and M3 gives $\chi^2=693.90$ for 472 DoF (BIC = 789.94).
In both sources, the broad-component model is preferred over the narrow-only alternatives.
This supports the inclusion of broad \ion{H}{$\beta$} emission and independently corroborates the broad-line interpretation inferred from \ion{H}{$\alpha$}.
The broad \ion{H}{$\beta$} component is fainter and therefore less well constrained, but its inclusion improves the description of the observed profile.

The [\ion{O}{III}] $\lambda\lambda4959,5007$ doublet was treated separately because its profiles show asymmetric structure that cannot be adequately described by a single narrow component.
The [\ion{O}{III}] $\lambda5007$ line is widely used to investigate narrow-line-region kinematics and ionized outflows, since its forbidden transition is not produced in the high-density broad-line region \citep[e.g.,][]{2014MNRAS.441.3306H,2018MNRAS.477.5115K}.
We therefore compared M1, with one Gaussian component per transition, M2, with two Gaussian components per transition, and M3, with three Gaussian components per transition.
Corresponding components of the two doublet lines were tied in wavelength separation and width, while their amplitudes were fitted independently, without imposing the theoretical $\lambda5007/\lambda4959$ flux ratio.
This approach was adopted to provide sufficient flexibility in reproducing the observed line profiles and to obtain reliable measurements of the total [\ion{O}{III}] fluxes, rather than to perform a detailed kinematic decomposition.
The fitting was performed iteratively, adding components successively to describe the asymmetric profiles.

\begin{figure*}[ht!]
    \centering

    \includegraphics[width=0.32\textwidth]{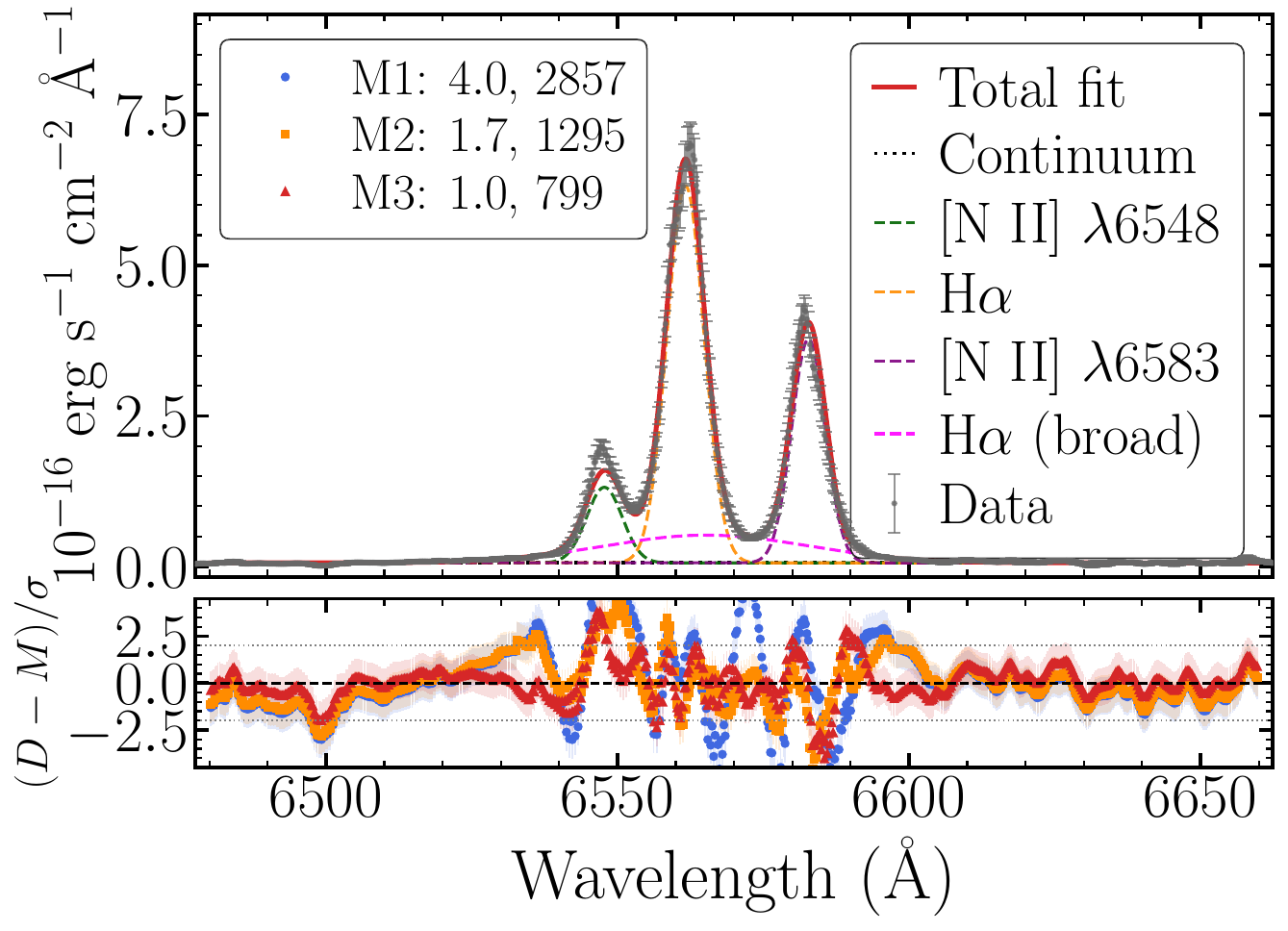}
    \includegraphics[width=0.32\textwidth]{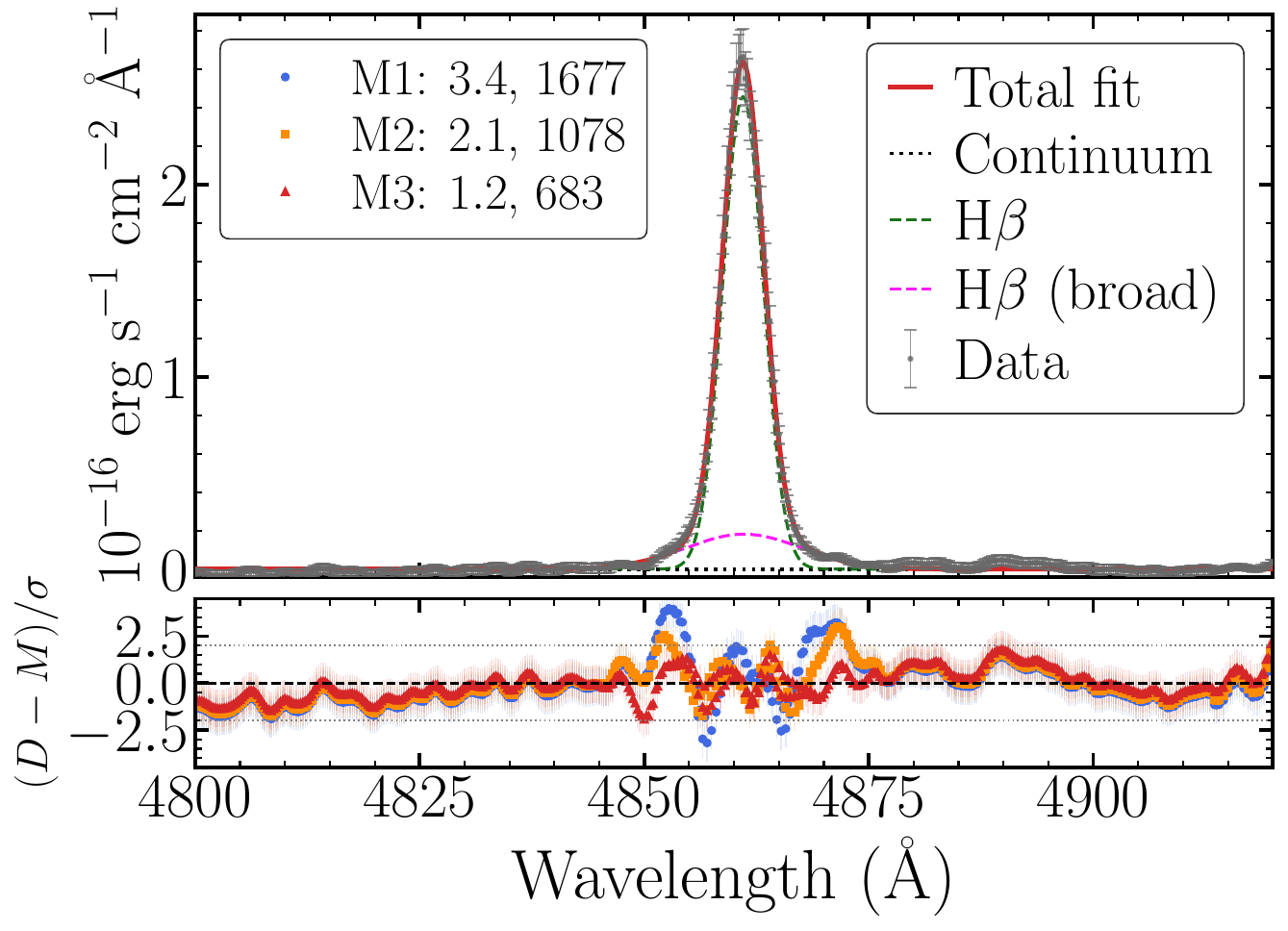}
    \includegraphics[width=0.32\textwidth]{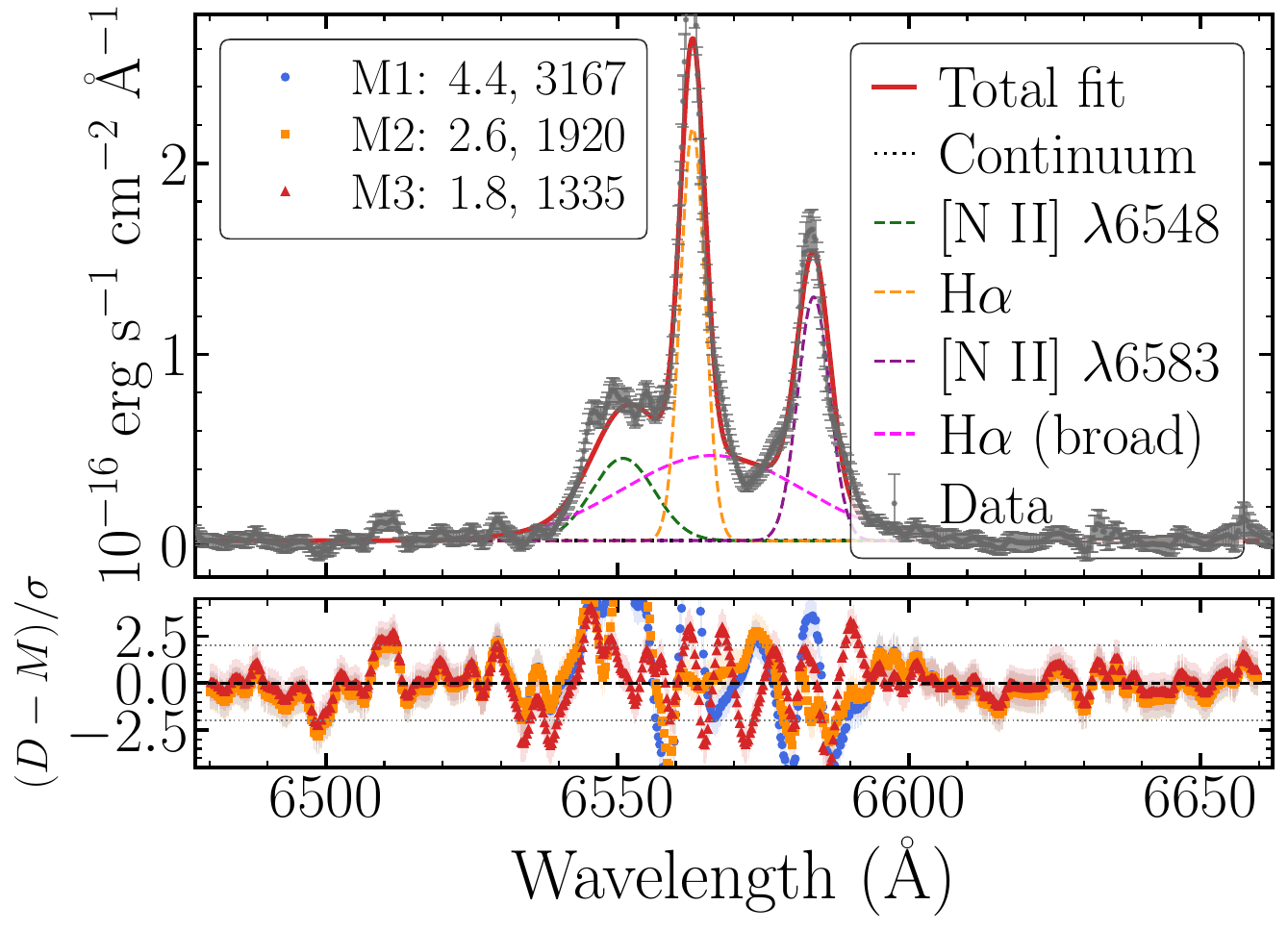}

    \vspace{0.3em}

    \includegraphics[width=0.32\textwidth]{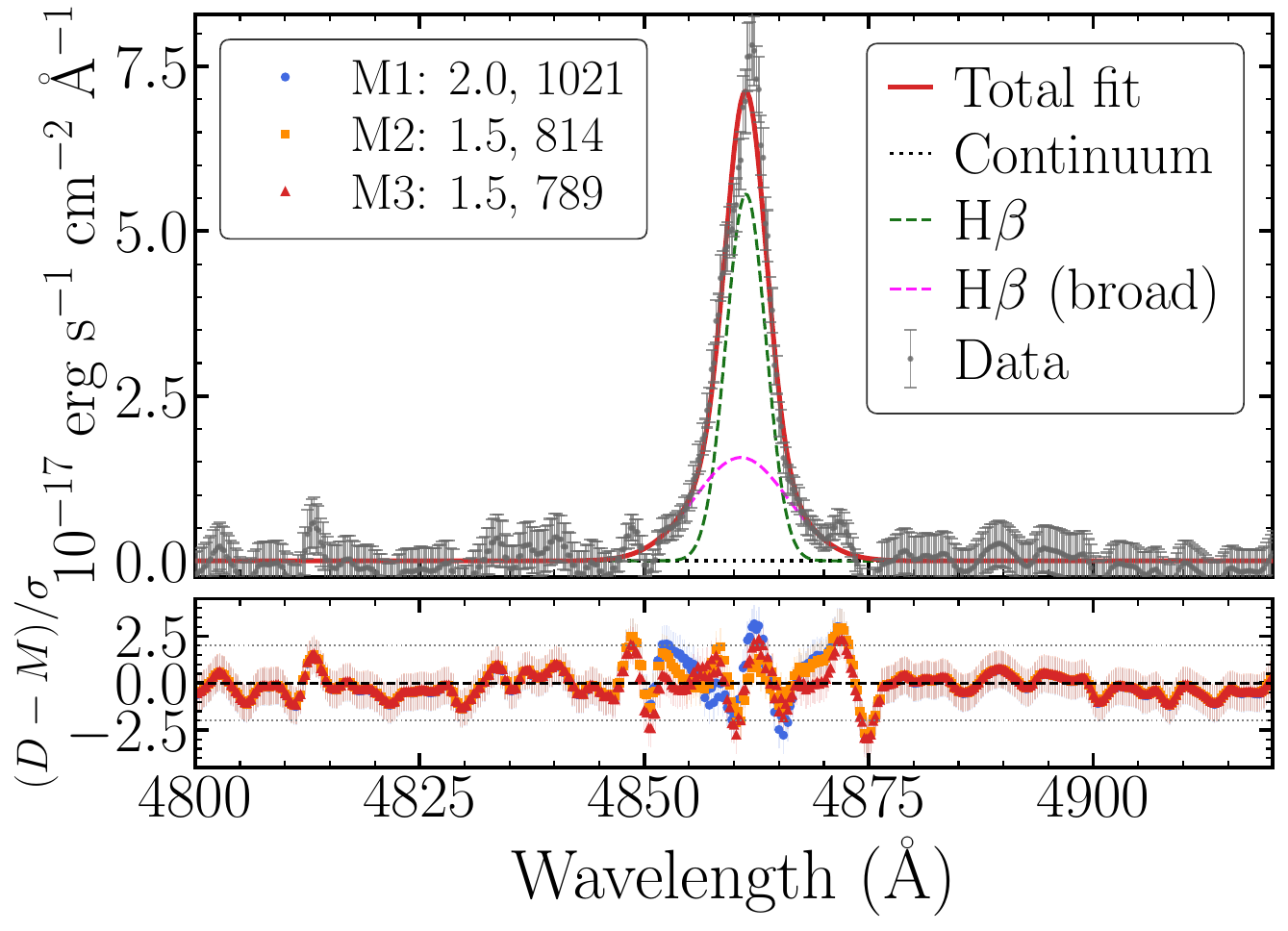}
    \includegraphics[width=0.32\textwidth]{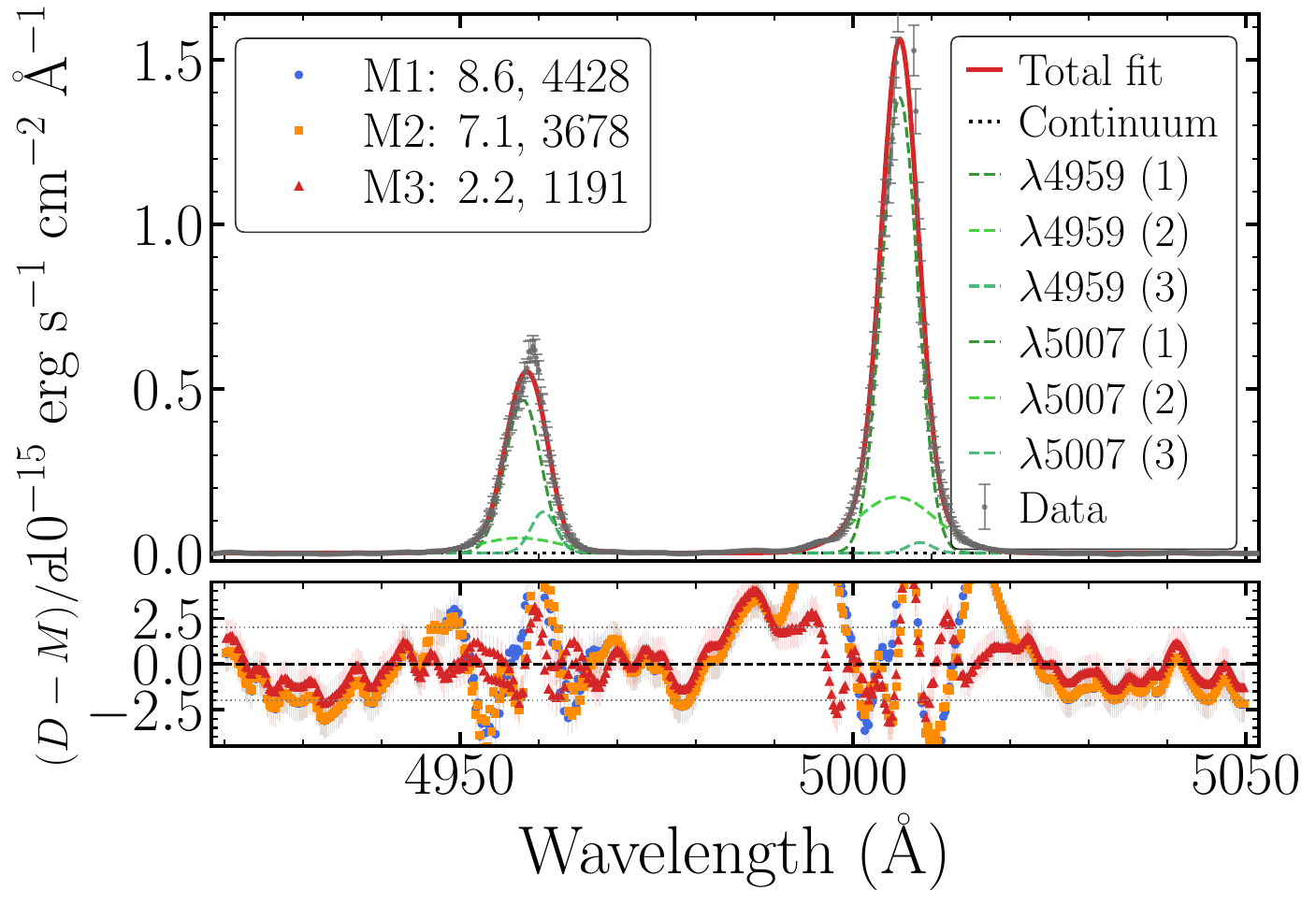}
    \includegraphics[width=0.32\textwidth]{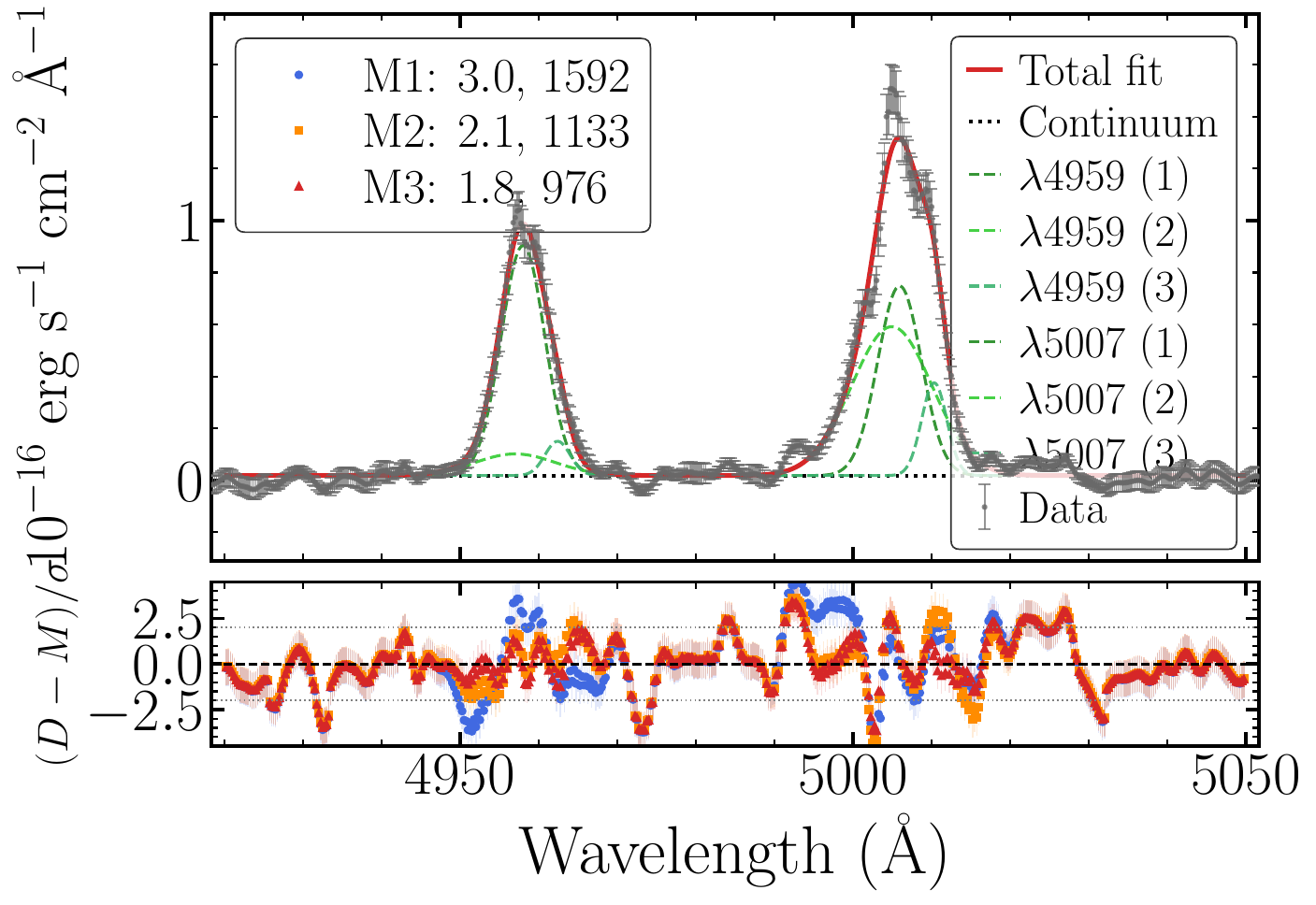}
    \vspace{0.3em}

    \includegraphics[width=0.32\textwidth]{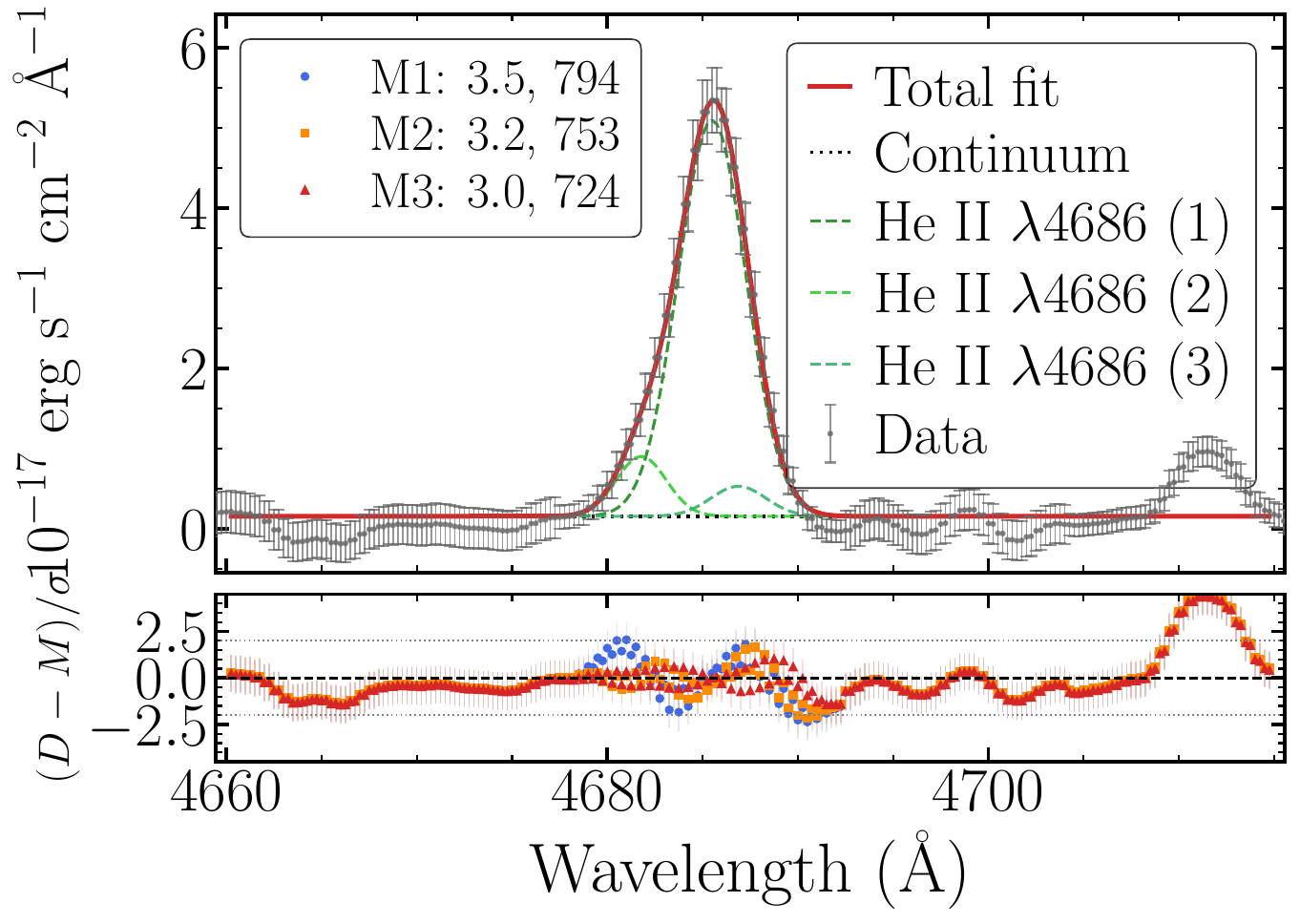}
    \includegraphics[width=0.32\textwidth]{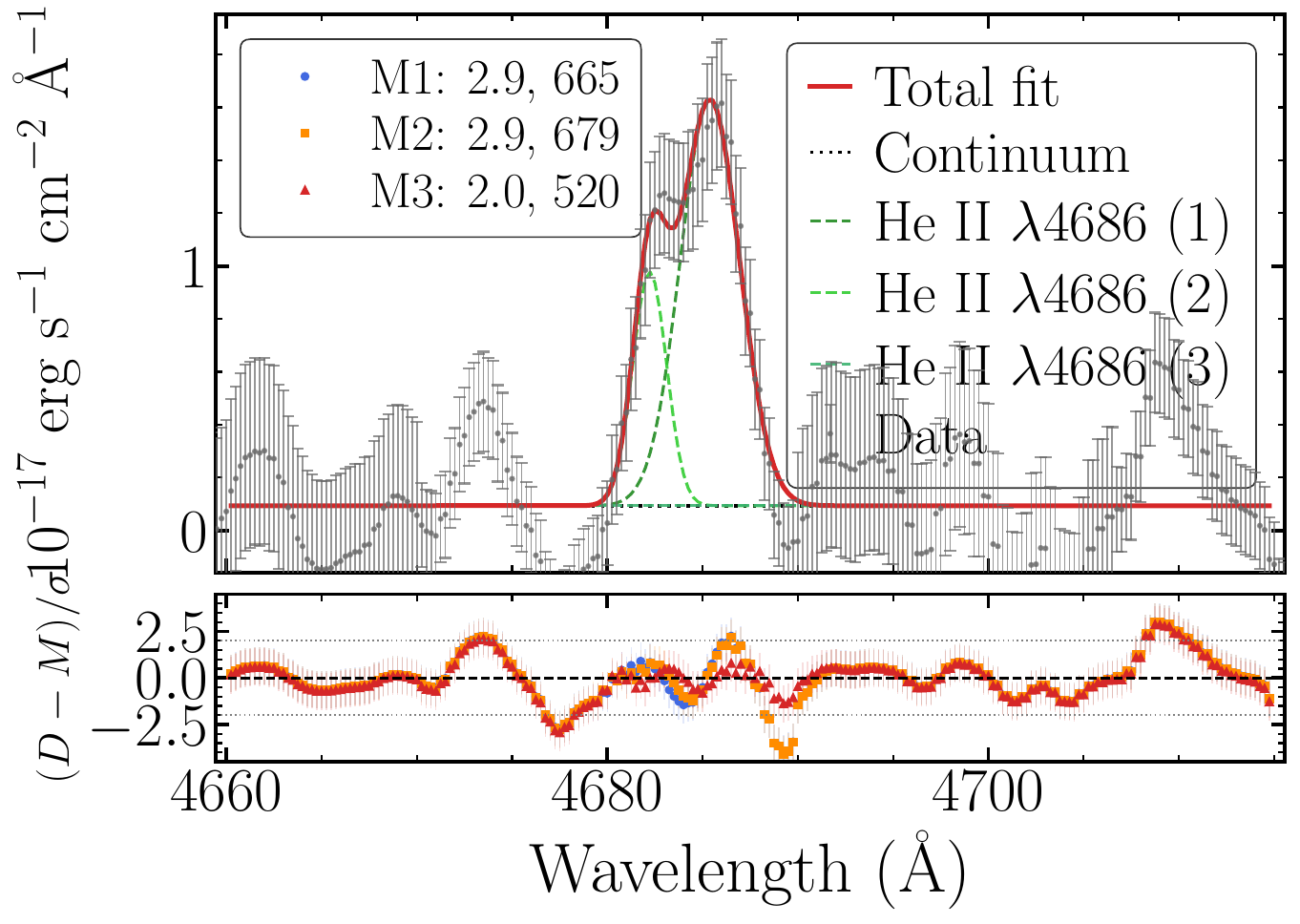}

    \caption{
        Multi-component Gaussian fits to the \ion{H}{$\alpha$}, \ion{H}{$\beta$}, [\ion{O}{III}], and \ion{He}{II} emission-line regions in the spectra of J1622+3521.
        The eight panels are arranged in three columns.
        From left to right and top to bottom, the panels show: \ion{H}{$\alpha$} (Source~1), \ion{H}{$\beta$} (Source~1), \ion{H}{$\alpha$} (Source~2), \ion{H}{$\beta$} (Source~2), [\ion{O}{III}] (Source~1), [\ion{O}{III}] (Source~2), \ion{He}{II} (Source~1), and \ion{He}{II} (Source~2).
        The red curve represents the total best-fit model, while the dotted black line marks the fitted continuum. Dashed curves show the individual Gaussian components.
        The \ion{H}{$\alpha$} region is modeled with narrow [\ion{N}{II}]~$\lambda\lambda6548,6583$, a narrow \ion{H}{$\alpha$} component, and a broad \ion{H}{$\alpha$} component.
        The \ion{H}{$\beta$} region is modeled using narrow and broad \ion{H}{$\beta$} Gaussian components.
        The [\ion{O}{III}] doublet is modeled with three narrow kinematic components per emission line, with the corresponding components tied in wavelength separation and width between the two transitions.
        The \ion{He}{II} $\lambda4686$ region is fitted with up to three Gaussian components to account for its potentially complex line profile.
        The lower panel in each spectrum shows the normalized residuals, $(\mathrm{Data}-\mathrm{Model})/\sigma$, for the three models (M1, M2, and M3), overlaid using different colors and marker symbols.
        The reduced $\chi^2$ and approximate Bayesian information criterion (BIC) values are reported in the legend for each model.
    }
    \label{fig:Balmer_fittings}
\end{figure*}

The [\ion{O}{III}] model comparison strongly favours M3 for both sources.
For Source~1, M1 gives $\chi^2=4396.52$ for 514 DoF (BIC = 4421.5), M2 gives $\chi^2=3621.27$ for 510 DoF (BIC = 3677.5), and M3 gives $\chi^2=1109.95$ for 506 DoF (BIC = 1191.2).
For Source~2, M1 gives $\chi^2=1560.62$ for 514 DoF (BIC = 1585.6), M2 gives $\chi^2=1077.04$ for 510 DoF (BIC = 1133.3), and M3 gives $\chi^2=894.56$ for 506 DoF (BIC = 975.8).
The substantial reductions in both $\chi^2$ and BIC indicate that three Gaussian components provide a much better description of the observed [\ion{O}{III}] profiles than the simpler alternatives.
Nevertheless, some residual structure remains, particularly around $\lambda5007$, so the fitted components should be regarded as a phenomenological description of the line profiles rather than unambiguous identifications of individual physical gas structures.
We also fitted the He~II $\lambda4686$ emission line using models with one, two, and three Gaussian components, and adopted the three-component model (M3), which provided the best fit according to the BIC.

\begin{figure*}[t]
    \centering
    \includegraphics[width=0.75\textwidth]{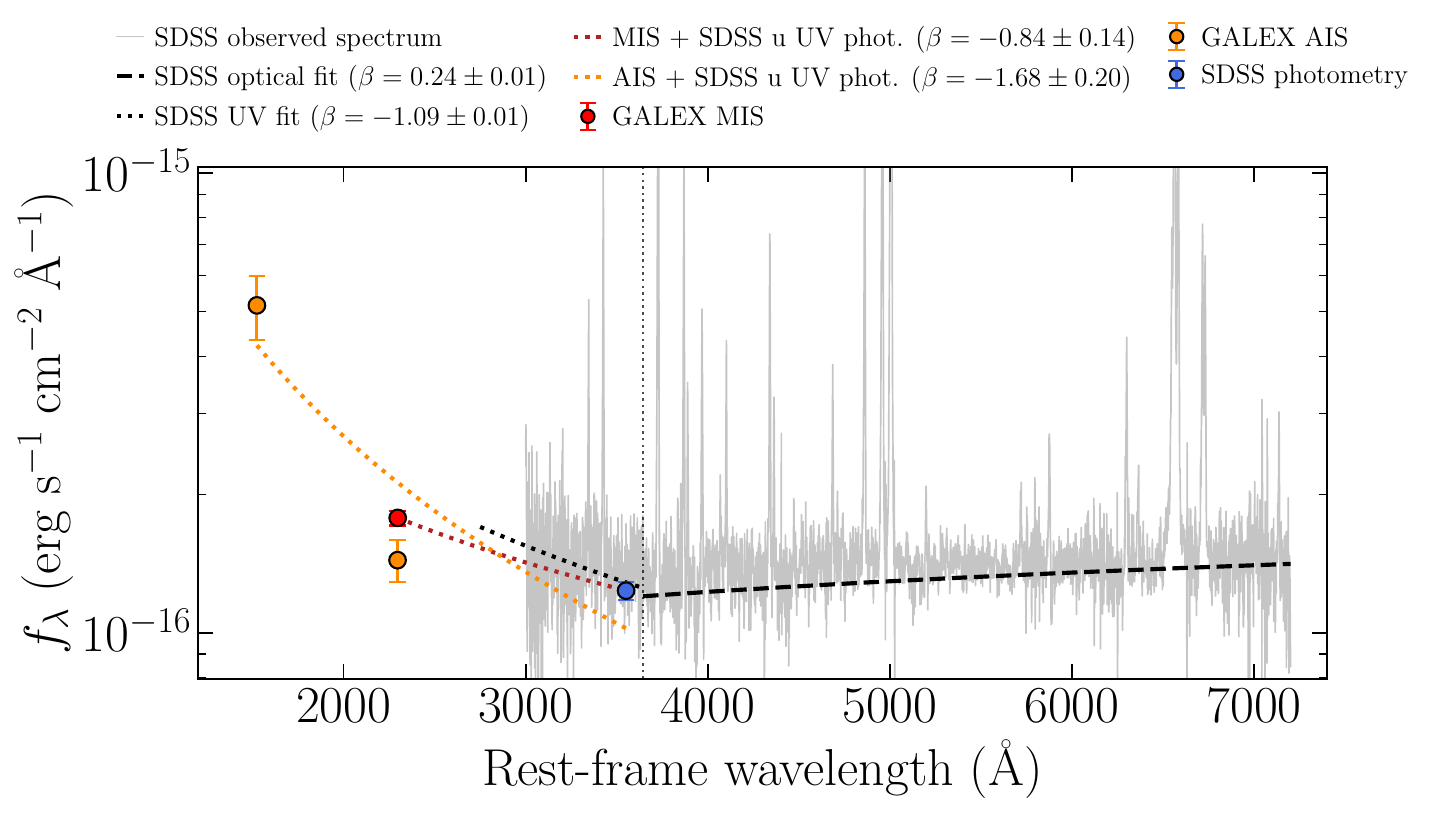}
    \caption{
        Rest-frame optical and UV continuum of J1622+3521.
        The grey curve shows the observed SDSS fibre spectrum, which includes contribution from both nuclei.
        The black dashed and dotted lines show power-law fits to the underlying stellar continuum modelled with \texttt{FADO}, redward and blueward of the Balmer break at $3645$~\AA, respectively.
        The red and orange points show the GALEX MIS and AIS measurements, respectively, while the blue point represents the SDSS $u'$ photometry.
        The red and orange dotted lines indicate power-law fits to the UV photometric measurements, combining the GALEX NUV (MIS) or FUV and NUV (AIS) measurements with the SDSS $u$-band point.
        }
    \label{fig:optical_uv_continuum}
\end{figure*}

Figure~\ref{fig:Balmer_fittings} presents the best-fitting M3 models for the four emission-line regions and the normalized residuals for M1, M2, and M3.
Overall, the model comparisons favour the inclusion of broad components in the Balmer lines and multiple narrow components in [\ion{O}{III}], providing a more complete description of the emission-line profiles in both sources.
Results of the fitting are provided in Table \ref{tab:balmer_results_compact}.

\subsubsection{Optical/UV continuum analysis}
\label{sec:optical_uv_continuum}

To further investigate the connection between J1622+3521 and the compact AGN populations identified at high redshift, we examined its optical and UV continuum shape.
We combined the SDSS optical spectrum with archival GALEX and SDSS photometry to construct the broadband UV-to-optical spectral energy distribution (SED) of the system (Figure~\ref{fig:optical_uv_continuum}).
The GALEX measurements were retrieved from the GR6/GR7 GALEX MCAT catalogue hosted by MAST \citep{2017A&A...603A..63G}, by querying within a $30''$ radius of the optical position of J1622+3521 and identifying counterparts in both the Medium Imaging Survey (MIS) and the All-sky Imaging Survey (AIS). The catalogue is available under DOI \href{https://doi.org/10.17909/T9H59D}{doi:10.17909/T9H59D}.
The MIS counterpart provides an NUV measurement, while the AIS counterpart provides both FUV and NUV photometry.
These measurements were complemented by the SDSS $ugriz$ photometry, with all photometric measurements representing the unresolved contribution from the two nuclei.
The observed AB magnitudes were converted into flux densities per unit wavelength using the corresponding effective wavelengths, and the photometric measurements were placed in the rest frame.
Furthermore, the broad-band optical photometry can be significantly affected by the strong emission lines present in the spectrum and therefore does not provide a reliable estimate of the underlying optical continuum slope.

We instead determined the optical continuum shape directly from the SDSS spectrum, which encompasses contributions from both nuclei within its fibre aperture.
We fitted power laws of the form $f_\lambda \propto \lambda^\beta$ to the underlying stellar continuum obtained from the \texttt{FADO} fit, considering separately the wavelength intervals blueward and redward of the Balmer break at $3645$~\AA.
The resulting slopes are $\beta_{\rm UV}=-1.09\pm0.01$ and $\beta_{\rm opt}=+0.25\pm0.005$, respectively, revealing a V-shaped continuum with a blue UV component and a mildly rising optical continuum.
The UV slope is independently supported by the GALEX photometry.
Using the MIS NUV measurement together with the SDSS $u$-band point, we obtain $\beta_{\rm UV}=-0.84\pm0.14$, while adopting the AIS FUV and NUV measurements together with the $u$-band point gives $\beta_{\rm UV}=-1.68\pm0.20$.
Although the two photometric estimates differ, both indicate a blue UV continuum and are broadly compatible with the slope measured from the SDSS spectrum.
The quoted uncertainties represent the formal fitting errors and do not necessarily account for all systematic uncertainties associated with the continuum decomposition and the non-simultaneous photometric measurements.
The optical slope is therefore based exclusively on the spectroscopic continuum, rather than on the potentially emission-line-contaminated broad-band photometry.

We do not use the Keck/DEIMOS spectra to derive independent continuum slopes for the two nuclei.
Although these observations spatially resolve the two components, their continua exhibit significant systematic uncertainties towards both the blue and red ends of the spectral coverage (see the 2D spectrum in Figure \ref{fig:Keck_speck}), making their overall spectral shapes unreliable for this purpose.
These limitations do not affect the emission-line analysis presented above, which relies on continuum-subtracted spectra and emission features sufficiently far from the affected spectral edges.
Consequently, the SDSS spectrum provides the most reliable estimate of the optical continuum shape, while the GALEX measurements independently constrain the unresolved UV emission.

The measured continuum slopes place the unresolved system within the UV--optical slope region commonly used to identify the V-shaped continua of LRDs \citep{2025ApJ...986..126K}.
In particular, the UV slope satisfies the adopted LRD criterion, while the positive optical slope is consistent with a redward-rising continuum, although the optical slope is relatively shallow.
However, the SDSS spectrum and the archival photometry are unresolved and include the combined contribution of both nuclei.
Consequently, the measured slopes characterize the integrated continuum of J1622+3521 and cannot establish whether either nucleus independently satisfies the continuum criteria.
We therefore describe J1622+3521 as a nearby system whose unresolved continuum falls within the LRD-like UV--optical slope region, without classifying the system or either nucleus as an LRD or LBD analog.
Spatially resolved continuum measurements, for example from future integral-field spectroscopy, will be required to assess the continuum properties of the individual nuclei.

\subsection{X-ray data \& source detection}
\label{sec:X_ray_source_detection}

The individual \textit{Chandra} observations were first reprocessed using the \texttt{chandra\_repro} task to ensure the application of the latest calibration files and screening criteria.
Before combining the datasets, the absolute astrometric alignment of each reprocessed event file was refined using the CIAO tools \texttt{wcs\_match} and \texttt{wcs\_update}, which register all observations to a common celestial reference frame.
Subsequently, each observation was reprojected to the tangent point of the first observation using the \texttt{reproject\_obs} task, ensuring a consistent world-coordinate system (WCS) across all files.
The reprojected event files were then merged into a single dataset by combining their photon lists with consistent WCS parameters.

The resulting merged event file was used to generate energy-filtered images in the soft (0.5--1.2~keV), medium (1.2--2~keV), hard (2--8~keV), and full (0.5--8~keV) bands using the \texttt{dmcopy} tool.
Exposure maps corresponding to each observation were also created and combined to account for variations in exposure time and instrumental response across the field.
An exposure-corrected flux image was then produced with the \texttt{fluximage} task, enabling the construction of images with uniform photometric scaling.
For visualization purposes, the merged event file was binned to a pixel scale of $0.123''$, corresponding to one-quarter of the native ACIS pixel size.
The resulting images are presented in Figure~\ref{fig:Chandra_PS1_images}.

Source detection was performed using CIAO’s \texttt{wavdetect} algorithm \citep{2002ApJS..138..185F}, which identifies statistically significant structures by correlating the X-ray image with a series of \textit{Mexican-hat} wavelets of increasing scale.
We applied \texttt{wavdetect} to the native-resolution ($0.492''$/pixel) image using standard wavelet scales.
The identified sources are shown in the bottom left panel of Figure~\ref{fig:Chandra_PS1_images}.
Their centroids are at 16:22:09.40, 35:21:07.2 (Source 1, west), and 16:22:09.56, 35:21:07.6 (Source 2, east).
The angular distance between the two detected X-ray centroids is $1.9''$, corresponding to a projected physical distance $d \simeq 8.4$~kpc.
The positions of the two optical centroids are spatially coincident with the two X-ray point sources identified in the \textit{Chandra} observations. 
However, for the spectral analysis, and to avoid overlap and cross-contamination from the \textit{Chandra} point-spread function (PSF), we adopt more compact circular apertures centered on the \texttt{wavdetect} positions with a radius of $1.19''$, and $0.78''$ for Sources 1 and 2, respectively
(Figure~\ref{fig:Chandra_PS1_images}, bottom right).

\subsection{X-ray spectral analysis}
\label{sec:X_spec_fit}

We extracted spectra for both targets separately from each of the three \textit{Chandra} observations using the \textsc{CIAO} \texttt{specextract} task. 
A single annular background region was constructed and used for both targets within all observations, centered at their midpoint with inner and outer radii of 7\,arcsec and 20\,arcsec, respectively. 
For each of the six corresponding spectra extracted, we used source\,$+$\,background circular regions for the West and East sources, respectively. 
We then used \texttt{specextract} to produce the on and off spectra with their corresponding response files after applying the appropriate exposure weighting and PSF corrections.

For both targets, we then separately combined their three epoch spectra after manually confirming that no significant intra-epoch variability had occurred. 
The \textsc{CIAO} \texttt{combine\_spectra} command was used to merge the spectra and generate the corresponding exposure- and response-weighted ARFs and RMFs.
The corresponding individual spectra generated for both targets were used for all subsequent analyses. Our primary aim from the X-ray spectral analysis was to derive robust intrinsic X-ray luminosities and line-of-sight column densities for both targets. 
A number of crucial considerations were required: (i) \textit{Blending:} the proximity of either target to one another meant that our fitting methodology needed to account for some level of blending, such that the X-ray spectra of both targets needed to be fit simultaneously to one another. (ii) \textit{Background:} the signal-to-noise ratio of either source was not equal, with the East source being sub-dominant to the background at soft X-ray energies. Thus proper consideration of the background was essential to account for its effects on the resulting parameter posteriors. (iii) \textit{Self-consistent X-ray spectral modelling:} the nature of the obscuration had to be considered with some levels of physical realism. 

The study of quasar and Seyfert AGN X-ray obscuration has been refined over decades of development \citep[see e.g.,][for a recent review]{Boorman24a_hexp}. 
However, the study of AGN obscuration within compact low-metallicity galaxies is considerably more novel and unexplored. Whilst low metallicity is known to have a significant impact on photoelectric absorption \citep{Wilman99}, there is currently very little understanding as to the connection between host galaxy metallicity and that of the circum-nuclear obscurer. Thus, our spectral fitting methodology needed to be sophisticated enough to consider low obscurer metallicity as a nuisance parameter such that our line-of-sight column densities and obscuration-corrected luminosities were as accurate as possible. For this purpose, we relied on the physically self-consistent spherical obscuration model of \citep{Brightman11}, which we refer to as \texttt{BNsphere}. With additional variable iron as well as other metal abundances as two separate parameters, \texttt{BNsphere} correctly accounts for Compton scattering, fluorescence, and photoelectric absorption occurring from a fully-covering sphere of obscuring gas to provide an estimated line-of-sight column density. We note that whilst a completely spherical obscurer is a difficult geometry to sustain in nature, it is the simplest possible geometry to assume in a physically self-consistent manner. 
Furthermore, the instigation of AGN obscuration during mergers is predicted to maximise the covering factor of the obscurer \citep[e.g.,][]{2017MNRAS.468.1273R,2018MNRAS.478.3056B,Ricci21}, and we note that such high covering factor obscurers are commonly invoked for LRDs, such as quasi-star models in which rapidly accreting black holes are embedded within dense, optically thick gaseous envelopes \citep{Rusakov26}.

Our X-ray spectral fits were performed within \textsc{PyXspec} -- the Python wrapper of \textsc{Xspec v12.12.1} \citep{Arnaud96,Gordon21} with the Poisson likelihood described by the C-statistic \citep{Cash79} and the Bayesian X-ray Analysis package \citep[BXA;][]{Buchner14,Buchner21} using the Python wrapper of the nested sampling package \textsc{MultiNest} \citep{Feroz09,Feroz19,Buchner14}. We use a sampling efficiency of 10$^{-3}$ with 10$^{4}$ live points within \textsc{MultiNest} as a means to optimise the robustness of the reconstructed parameter posteriors \citep{Dittmann24}. The spectral fit included fixed Galactic absorption along the line-of-sight to both sources with $N_{\rm H,\,Gal}$\,=\,1.03\,$\times$\,10$^{20}$\,cm$^{-2}$ \citep{Willingale13} using the \texttt{TBabs} model and abundances from \citep{Wilms00}. All spectral fitting parameter estimates are listed as the mode of their respective posterior distributions with uncertainties quoted as the 90\% Highest Density Interval.

\subsubsection{Free Parameters}\label{subsubsec:freepars}

To incorporate as much information as possible from the observed data in our fits, we include no binning and instead rely on a background model to avoid systematic fitting biases \citep[e.g.,][]{Buchner23}. 
For convenience, we choose to use the automated background models that are publicly accessible through BXA. 
Background model templates were generated by fitting large samples of background observations for a variety of instruments using Principal Component Analysis (PCA) by \citet{Simmonds18}. 
The automated background fitter within BXA then takes the PCA models of \citet{Simmonds18} and performs a bespoke fit for each of the three X-ray background spectra we analyse in this work, making sure to include additional Gaussian emission lines if statistically required. 
We visually verify for each spectrum that the background model provides an acceptable fit to the observed background spectra before incorporating the model into our X-ray spectral fits. 
Since the PCA components that make up each background model do not necessarily represent physical components, it would not make physical sense to let those components all vary freely during the fit. 
However, to include some level of realistic noise arising from the imperfect knowledge of the background, we do allow the overall normalisation of each full background model to vary as a free parameter along with the spectral parameters of each fit. 
The background scaling normalisation is allowed to vary with a log-uniform prior over two orders of magnitude centered at the expected value.

We assume the same X-ray spectral model for both sources, with two additional pre-multiplying `mixing' constants to account for the contaminating flux from Source~1 in the Source~2 spectrum and vice versa. 
Each source spectrum comprises the AGN emission represented with the \texttt{BNsphere} model and a soft X-ray power-law designed to reproduce any residual flux that arises from unresolved X-ray binaries in either host galaxy and/or AGN primary emission leaking through small gaps in the primary obscurer.
Per source, the \texttt{BNsphere} model has variable photon index and normalisation of the primary coronal continuum, which we assign a Gaussian prior of 1.80\,$\pm$\,0.15 which is a conservative assumption consistent with previous AGN studies \citep[e.g.,][]{Ricci17_bassV}, designed to avoid artificially hard spectral shapes \citep[c.f.][]{2025ApJ...978..118B} and a log-uniform prior between 10$^{-10}$\,--\,1, respectively. 
The line-of-sight column density of the obscurer was assigned a log-uniform prior between 10$^{20}$\,--\,10$^{26}$\,cm$^{-2}$, and its elemental composition was assigned log-uniform priors between 0.1--1 Solar metallicity for the iron abundance and joint abundance of all other metals. 
The contaminating soft X-ray component was allowed to vary in normalisation with a log-uniform prior, but was forced to never exceed 10\% of the primary AGN continuum. 
Its photon index was also allowed to vary with a uniform prior between 1\,--\,10. Considering each source had a total of nine free parameters, including contamination mixing and background normalisation, our full model was described by 18 total free parameters.



\subsection{IR photometry}
\label{sec:IR_photometry}

Additionally, we examine the infrared properties of J1622+3521 as observed with the Wide-field Infrared Survey Explorer \citep[\textit{WISE};][]{2010AJ....140.1868W}.  
J1622+3521 appears unresolved due to the relatively large point spread function of the \textit{WISE} bands, which is $\geq 5.6''$ in W1 and increases to $\sim 12''$ in W4.  
We adopt the photometry reported in the AllWISE Source Catalog \citep{2010AJ....140.1868W}, available through \href{https://irsa.ipac.caltech.edu/data/WISE/docs/release/AllWISE/}{IRSA} under DOI \href{https://doi.org/10.26131/IRSA1}{10.26131/IRSA1}.
J1622+3521 is red in the near- and mid-IR bands covered by \textit{WISE}, with $\mathrm{W1-W2} = 1.38 \pm 0.03$\,mag and $\mathrm{W2-W3} = 3.05 \pm 0.03$\,mag.  
These colors place the source well within the widely used AGN selection wedges \citep{2011ApJ...735..112J,2012ApJ...753...30S,2012MNRAS.426.3271M,2018ApJ...858...38S}.  
Moreover, the \textit{WISE} colors of J1622+3521 are consistent with those of the majority of IR-detected Green Pea galaxies, which, however, were not classified as AGN based on their optical spectra \citep[e.g.,][]{2016ApJ...832..119H,2024A&A...688A.159K}.

\section{Results}
\label{sec:results}

\subsection{X-ray Spectral Fitting Results}\label{subsec:xrayres}

\begin{figure*}[ht!]
    \centering
    \includegraphics[width=\linewidth]{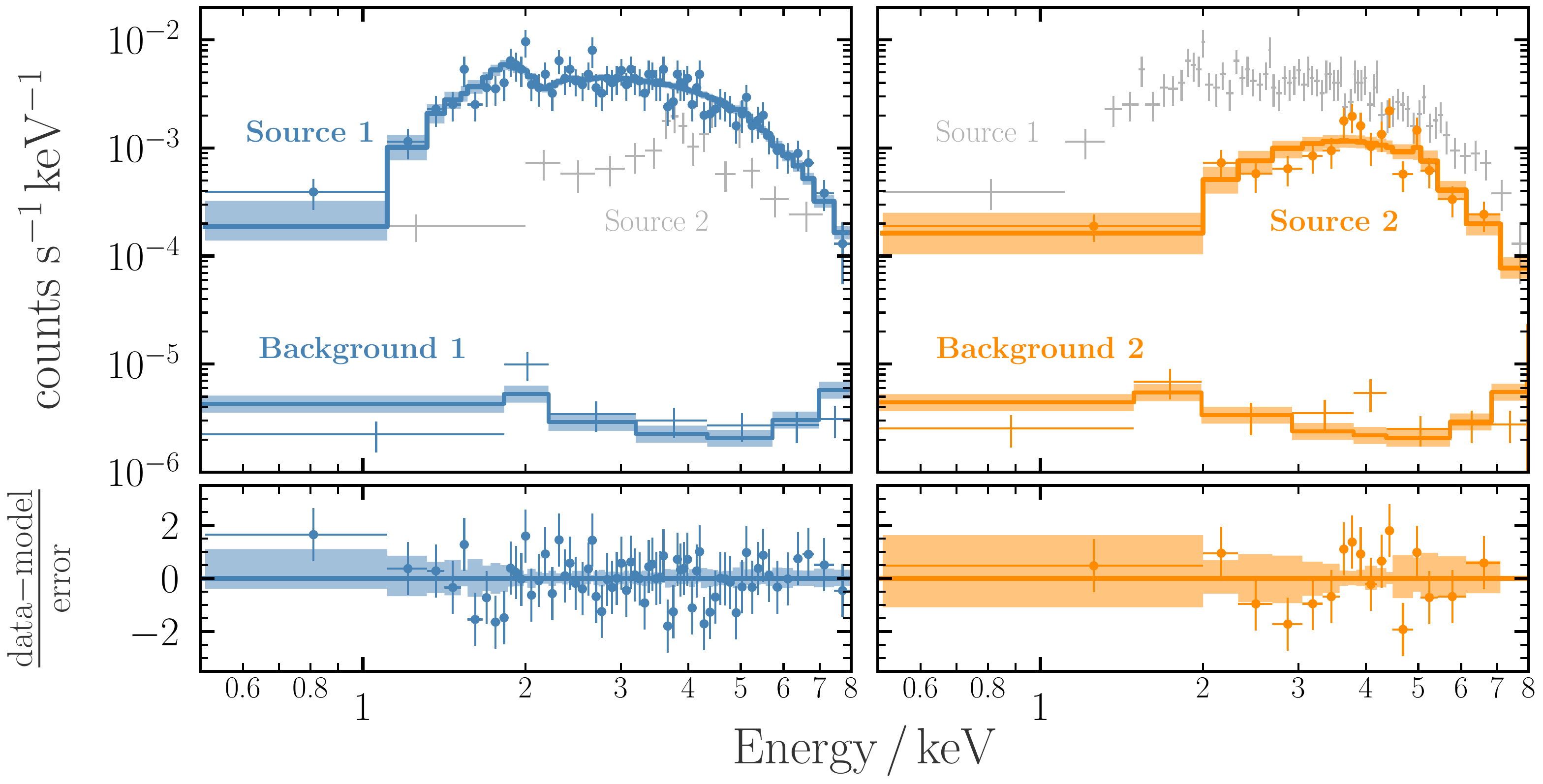}
    \caption{
    Left and right panels present the X-ray spectral posterior constraints for Source~1 and~2 in blue and orange, respectively. In both panels, the corresponding background spectrum and fit relevant to the source considered is shown in the bottom portion of the spectral panels. The lower portion of the left and right panels presents the residuals of the source\,$+$\,background data with error bars, as well as the posterior range around the median with shaded regions.
    }
    \label{fig:X_ray_spectra}
\end{figure*}

The joint fit to the \textit{Chandra} spectra of Source~1 and~2 is shown in Figure~\ref{fig:X_ray_spectra}, in which the model is able to simultaneously explain the observed source\,$+$\,background and background spectra well. 
We find a logarithmic line-of-sight column density for Source~1 and~2 of log\,($N_{\rm H}$\,/\,cm$^{-2}$)\,=\,$22.8_{-0.2}^{+0.5}$ and $23.6_{-0.5}^{+0.4}$, respectively. 
The corresponding logarithmic intrinsic 2\,--\,10\,keV luminosities for Source~1 and~2 are log\,($L_{2-10\,{\rm keV}}$\,/\,erg\,s$^{-1}$)\,=\,$43.93_{-0.05}^{+0.04}$ and $43.55$\,$\pm$\,0.11, respectively. We note that the general effect of including variable metallicities for either source as nuisance parameters has increased the level of uncertainty for column density that is self-consistently included in the ultimate estimate of intrinsic luminosity, as required. Lastly, we find intrinsic coronal photon index estimates of $1.76_{-0.21}^{+0.20}$ and $1.79_{-0.23}^{+0.25}$ for Source~1 and~2, respectively. 
The broad consistency with the input prior of 1.80\,$\pm$\,0.15 is clear, suggesting that the measured spectra are insensitive to the parameter under the adopted model. 
A plausible reason is the result of competing parameters describing the overall spectral shape of either source (e.g., metallicity, line-of-sight column density, contamination mixing fractions). Given the substantial luminosity of both sources above 2\,keV and their overall spectral shapes, our X-ray spectral fitting unambiguously confirms their AGN nature.

\subsection{Optical emission-line diagnostics}
\label{sec:BPT}

\begin{figure}
    \centering
    \includegraphics[width=\columnwidth]{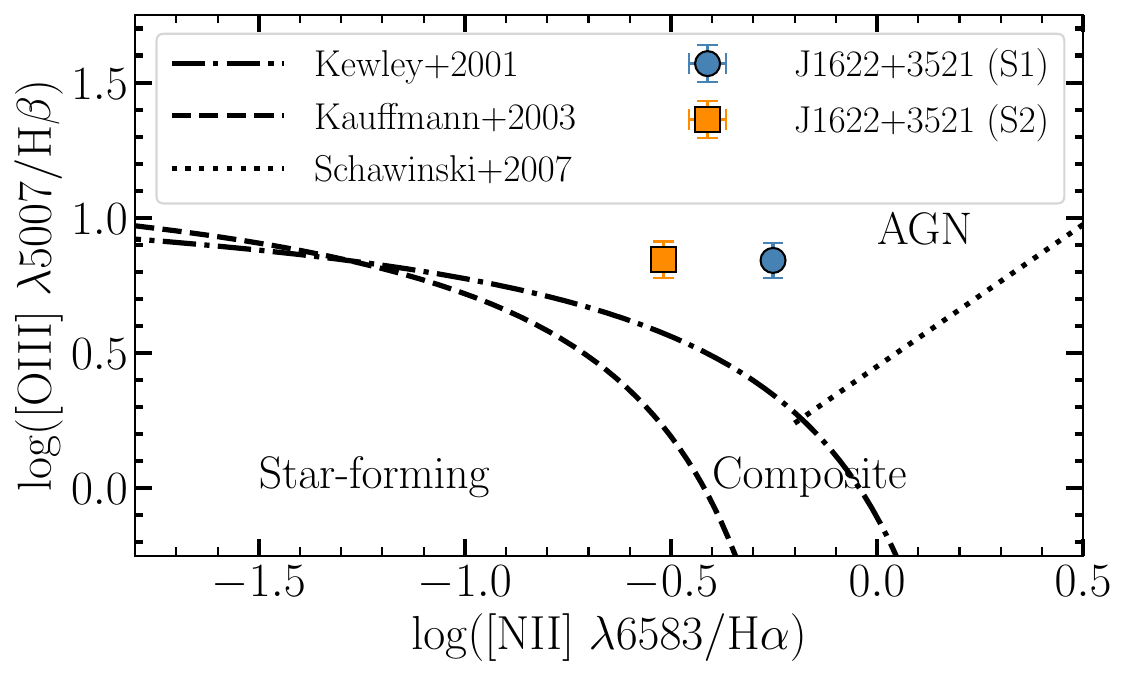}
    \caption{
    BPT diagnostic diagram for J1622+3521 based on the narrow emission-line ratios derived from the spectral decomposition. 
    The blue circle and orange square correspond to the DEIMOS measurements of Source~1 and Source~2, respectively.
    The dashed, dash-dotted, and dotted curves show the classification boundaries from \citep{2003MNRAS.346.1055K}, \citep{2001ApJ...556..121K}, and \citep{2007MNRAS.382.1415S}, respectively. 
    }
    \label{fig:BPT}
\end{figure}

To further investigate the dominant ionization mechanism in the two nuclei, we constructed a Baldwin--Phillips--Terlevich (BPT) diagnostic diagram \citep{1981PASP...93....5B} using the narrow emission-line ratios derived from the spectral decomposition (Table \ref{tab:balmer_results_compact}). 
Figure~\ref{fig:BPT} shows the locations of Source~1 and Source~2 in the $\log(f_{\rm [OIII]~\lambda5007}/f_{\mathrm{H}\beta})$ versus $\log(f_{\rm [NII]~\lambda6583}/f_{\mathrm{H}\alpha})$ plane. 
Both measurements lie securely within the AGN region of the diagram \citep{2001ApJ...556..121K,2003MNRAS.346.1055K,2007MNRAS.382.1415S}, confirming that the optical line emission is dominated by photoionization from accreting supermassive black holes rather than by star formation alone. 
The two DEIMOS nuclei occupy very similar positions, with high excitation ratios of $\log([\ion{O}{III}]/\mathrm{H}\beta)\sim0.9$.

Therefore, the BPT classification independently supports the dual-AGN interpretation inferred from the X-ray luminosities and broad Balmer emission.
Combined with the luminous hard X-ray emission and signatures from a broad-line region, the optical narrow-line ratios establish J1622+3521 as a bona fide dual AGN system hosted within a compact Green Pea galaxy.
This interpretation is further reinforced by the detection of high-ionization (coronal) emission lines, including [\ion{Fe}{VII}]~$\lambda6087$ in the DEIMOS spectra and [\ion{Ne}{V}]~$\lambda\lambda3346,3426$ in the SDSS fibre spectrum (Appendix \ref{app:high_ionization}).
The production of these ions requires photons with energies of $\sim99$ and $\sim97$~eV, respectively, well above those typically produced by stellar populations, and therefore, they are widely regarded as robust signatures of AGN photoionization and actively accreting supermassive black holes \citep[e.g.,][]{2006MNRAS.371.1559G,2010MNRAS.405.1315M,2024ApJ...976..130M}.

\subsection{Broad-line emission and black hole mass estimates}
\label{sec:BH_mass}

We estimate supermassive black hole masses using single-epoch virial scaling relations based on the broad \ion{H}{$\alpha$} emission \citep[e.g.,][]{2005ApJ...630..122G,2007ApJ...670...92G,2013ApJ...775..116R}.
Throughout this work, we adopt the calibration of \citet{2013ApJ...775..116R}, which is optimized for low-mass broad-line AGN systems similar to J1622+3521.

To estimate the extinction toward the broad-line region (BLR), we use the broad \ion{H}{$\alpha$}/\ion{H}{$\beta$} Balmer decrements, assuming an intrinsic ratio of 3.1 commonly adopted for broad-line AGN \citep[e.g.,][]{2005ApJ...630..122G,2008MNRAS.383..581D,2019MNRAS.483.1722L} and adopting the attenuation law of \citet{2000ApJ...533..682C}.
This yields $E(B-V)_{\rm BLR}=0.687\pm0.121$ for Source~1 and $0.854\pm0.389$ for Source~2.
The corresponding attenuations at \ion{H}{$\alpha$} are $A({\rm H}\alpha)=1.742$ and $2.166$ mag, respectively.

The intrinsic BLR Balmer decrement remains somewhat uncertain, with values of $\sim2.7$ also reported in the literature \citep[e.g.,][]{2017MNRAS.467..226G}.
Using such values would lead to different extinction corrections but would not alter the main conclusions of this work.

For comparison, the narrow-line Balmer decrements imply $E(B-V)_{\rm NLR}\simeq0.19$ and $\simeq1.01$ for Sources~1 and 2, respectively.
The substantially larger BLR reddening in Source~1 suggests additional obscuration toward the nuclear region, whereas the broad- and narrow-line extinctions are broadly consistent within the uncertainties in Source~2.
Such differences between BLR and NLR reddening are commonly observed in AGN and indicate that the two emitting regions need not experience the same line-of-sight extinction \citep[e.g.,][]{2019MNRAS.483.1722L}.

The observed broad \ion{H}{$\alpha$} luminosities are $L({\rm H}\alpha_{\rm broad})=4.780~(\pm0.290)\times10^{41}\,{\rm erg\,s^{-1}}$ for Source~1 and $4.038~(\pm0.111)\times10^{41}\,{\rm erg\,s^{-1}}$ for Source~2.
After correcting for extinction, we obtain $L({\rm H}\alpha_{\rm broad})=2.378~(\pm0.689)\times10^{42}\,{\rm erg\,s^{-1}}$ and $2.968~(\pm2.698)\times10^{42}\,{\rm erg\,s^{-1}}$, respectively.
The broad-line FWHM measurements are $1894\pm56\,{\rm km\,s^{-1}}$ for Source~1 and $1655\pm36\,{\rm km\,s^{-1}}$ for Source~2.

Using the extinction-corrected broad \ion{H}{$\alpha$} luminosities and broad-line FWHM measurements, we derive the following black hole masses with the adopted \citet{2013ApJ...775..116R} calibration:
\begin{itemize}
    \item Source~1: $\log(M_{\rm BH}/M_\odot)=7.32\pm0.06$,
    \item Source~2: $\log(M_{\rm BH}/M_\odot)=7.24\pm0.19$.
\end{itemize}
For comparison, the \citet{2007ApJ...670...92G} calibration yields $M_{\rm BH}=1.20~(\pm0.24)\times10^7\,M_\odot$ for Source~1 and $1.02~(\pm0.57)\times10^7\,M_\odot$ for Source~2.
These estimates are consistent with the adopted masses and do not alter the main conclusions of this work.

The quoted uncertainties reflect the propagated statistical uncertainties from the luminosity and FWHM measurements only and do not include the intrinsic systematic scatter of $\sim0.4$--0.5 dex associated with single-epoch virial supermassive black hole mass estimators.

We also measured stellar velocity dispersions from the optical spectra fits with \texttt{FADO}.
However, we do not adopt black hole masses derived from the local $M_{\rm BH}$--$\sigma_*$ relation.
In compact interacting systems such as J1622+3521, the observed stellar kinematics may be influenced by rotation, tidal motions, and non-equilibrium dynamical effects associated with the merger, such that the measured velocity dispersions may not represent the relaxed bulge kinematics assumed by local $M_{\rm BH}$--$\sigma_*$ calibrations \citep[e.g.,][]{2013ARA&A..51..511K,2014ApJ...786...12S}.
We therefore adopt the broad \ion{H}{$\alpha$}-based virial black hole masses throughout this work.

    \subsection{Accretion properties and Eddington ratios}
\label{sec:eddington}

Using the intrinsic X-ray luminosities derived from the BXA spectral modeling together with the virial supermassive black hole masses inferred from the broad \ion{H}{$\alpha$} emission, we estimate the accretion properties of both nuclei.
The absorption-corrected 2--10\,keV luminosities are
$\log (L_{\mathrm{2-10\,keV},1}/{\rm erg~s^{-1}})=43.93_{-0.05}^{+0.04}$ and
$\log (L_{\mathrm{2-10\,keV},2}/{\rm erg~s^{-1}})=43.55\pm0.11$ for Source~1 and Source~2, respectively.

Adopting a bolometric correction of $k_{\rm bol}=20$, appropriate for rapidly accreting NLS1-like systems \citep[e.g.,][]{2007MNRAS.381.1235V,2009MNRAS.392.1124V,2020A&A...636A..73D}, we derive:
\begin{align}
\log (L_{\rm bol,1}/{\rm erg~s^{-1}}) &= 45.23 \pm 0.02,\\
\log (L_{\rm bol,2}/{\rm erg~s^{-1}}) &= 44.85 \pm 0.06.
\end{align}

Combining these values with the virial black hole mass estimates of $\log(M_{\rm BH,1}/M_\odot)=7.32\pm0.06$ and $\log(M_{\rm BH,2}/M_\odot)=7.24\pm0.19$ yields Eddington ratios of:
\begin{align}
\log (\lambda_{\rm Edd,1}) &= -0.19 \pm 0.06,\\
\log (\lambda_{\rm Edd,2}) &= -0.49 \pm 0.20,
\end{align}
corresponding to $\lambda_{\rm Edd}\simeq0.65$ and $0.32$ for Source~1 and Source~2, respectively.
The quoted uncertainties reflect the propagated statistical uncertainties from the virial mass estimates and X-ray luminosities, while the true systematic uncertainties are likely larger owing to the intrinsic scatter associated with single-epoch virial black hole mass estimators and bolometric corrections.
Both nuclei are therefore accreting efficiently, with Source~1 exhibiting a particularly high accretion rate characteristic of rapidly growing Seyfert~1 and NLS1-like systems \citep[e.g.][]{2016MNRAS.462.1256C}.

The adopted bolometric correction is broadly consistent with values inferred for local Seyfert~1 and NLS1-like AGN.
Reasonable variations in $k_{\rm bol}$ within the range commonly reported for such systems ($k_{\rm bol}\sim10$--30) would shift the inferred Eddington ratios by approximately $-0.30$ to $+0.18$\,dex relative to our adopted value of $k_{\rm bol}=20$ and would not affect the main conclusions.

As an independent check on the bolometric luminosities, we estimate $L_{\rm bol}$ from the extinction-corrected broad \ion{H}{$\alpha$} luminosities using the empirical relation of \citet{2012MNRAS.423..600S}, adopting $L_{\rm bol}\simeq130\,L_{\rm broad,H\alpha}$.
Using the updated extinction-corrected broad \ion{H}{$\alpha$} luminosities we obtain:
\begin{align}
\log (L_{\rm bol,H\alpha,1}/{\rm erg~s^{-1}}) &= 44.49 \pm 0.13,\\
\log (L_{\rm bol,H\alpha,2}/{\rm erg~s^{-1}}) &= 44.59 \pm 0.39.
\end{align}
These estimates are lower than the X-ray-based bolometric luminosities by approximately $0.74$\,dex for Source~1 and $0.26$\,dex for Source~2.
The difference is more pronounced for Source~1, whereas the estimates for Source~2 are broadly consistent given the large uncertainty in its extinction-corrected broad-line luminosity.
These differences may reflect the intrinsic scatter in the broad-line-to-bolometric conversion, uncertainties in the BLR extinction correction, and variations in the spectral energy distributions of the two nuclei.
We therefore retain the X-ray-based bolometric luminosities for our Eddington-ratio estimates and use the broad-line-based values as an independent consistency check.

\subsection{Comparison between He~II and X-ray emission}
\label{sec:HeII_Xray}

To further investigate the connection between the ionizing radiation field and the accretion properties of the two nuclei, we compare the observed \ion{He}{II}~$\lambda4686$ luminosities with the observed soft X-ray luminosities in the 0.5--2\,keV band. 
The \ion{He}{II} emission line traces photons with ionization energies above 54.4\,eV and is therefore sensitive to the presence of hard ionizing radiation. 
Although nebular \ion{He}{II} emission can arise from several mechanisms, including AGN activity, shocks, X-ray binaries, and very young stellar populations (e.g., Wolf-Rayet stars), the observed narrow \ion{He}{II}/\ion{H}{$\beta$} ratios of $\sim0.19$ and $\sim0.25$ for Source~1 and Source~2 place both nuclei firmly within the AGN-dominated regime of the diagnostic diagrams presented by \citet{2012MNRAS.421.1043S}.
Together with the detection of the high-ionization lines [\ion{Ne}{V}] and [\ion{Fe}{VII}], this suggests that stellar photoionization alone is unlikely to generate the hard ionizing continuum required to produce the observed \ion{He}{II} emission.

\begin{figure}
    \centering
    \includegraphics[width=\columnwidth]{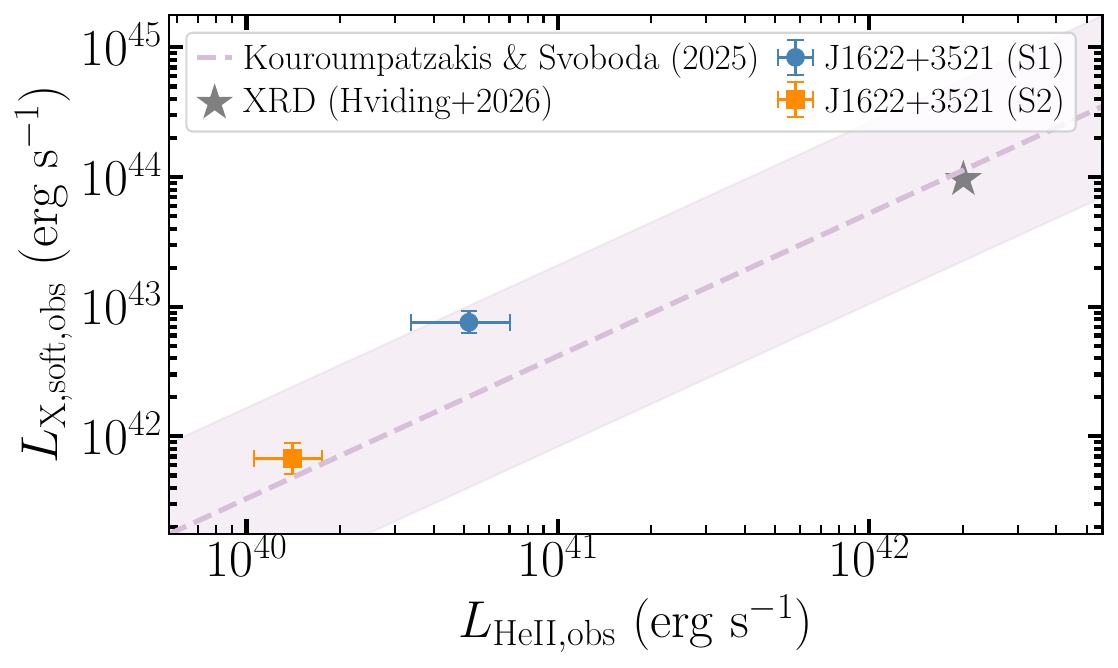}
    \caption{
    Comparison between the observed narrow \ion{He}{II}~$\lambda4686$ luminosity and the observed soft X-ray luminosity ($0.5$--$2$\,keV) for the two nuclei of J1622+3521.
    The dashed line and shaded region show the empirical AGN relation and intrinsic scatter derived by \citet{2025A&A...696A.133K}. 
    Blue and orange symbols correspond to Source~1 and Source~2, respectively. 
    The gray star symbol marks the "X-ray Dot" from \citet{2026ApJ..1000L..18H}. 
    }
    \label{fig:HeII_Xray}
\end{figure}

The measurements for Source~1 and Source~2 are compared with the empirical AGN relation derived by \citet{2025A&A...696A.133K} in Figure~\ref{fig:HeII_Xray}. 
Both nuclei lie very close to the empirical \ion{He}{II}--soft X-ray relation and remain fully consistent with its intrinsic scatter.
We additionally compare the system with the recently discovered X-ray-detected Red Dot (XRD) presented by \citet{2026ApJ..1000L..18H}. 
The XRD occupies the high-luminosity extension of the same relation and lies broadly consistent with the AGN sequence traced by J1622+3521.
The agreement between the X-ray Dot and J1622+3521 suggests that compact, rapidly accreting supermassive black holes may follow a common ionizing-continuum scaling relation across several orders of magnitude in luminosity and a wide range of cosmic epochs.

The consistency between the soft X-ray and \ion{He}{II} luminosities provides independent evidence that J1622+3521 hosts two actively accreting supermassive black holes embedded within compact, highly ionized environments.
More generally, this comparison provides additional support for using narrow \ion{He}{II} emission as a proxy for intrinsic AGN power in compact and partially obscured systems where direct X-ray measurements or broad-line diagnostics may be difficult to obtain.


\section{Discussion}
\label{sec:Discussion}

\subsection{Comparison with local AGN and dual-AGN populations}
\label{sec:Comp_with_local_and_dual_AGN}

\begin{figure}
    \centering
    \includegraphics[width=\columnwidth]{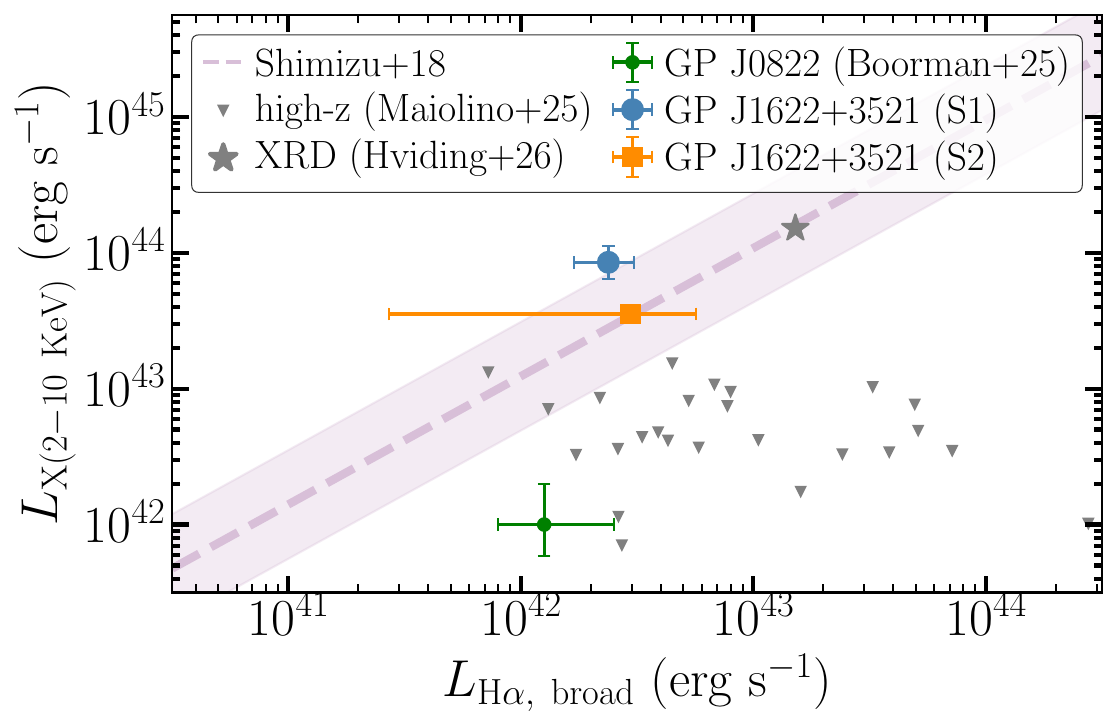}
    \caption{
    Comparison between the extinction-corrected broad \ion{H}{$\alpha$} luminosity and the intrinsic 2--10\,keV X-ray luminosity for J1622+3521. 
    Blue and orange points correspond to Source~1 and Source~2, respectively. 
    The dashed line and shaded region show the empirical AGN relation of \citet{2018ApJ...856..154S} and its intrinsic scatter. 
    The green point marks the Green Pea galaxy~J0822+2241 from \citet{2025ApJ...988..157B}, while gray triangles indicate the X-ray upper limits for compact high-redshift AGN reported by \citet{2025MNRAS.538.1921M}.
    }
    \label{fig:shimizu}
\end{figure}

The intrinsic broad \ion{H}{$\alpha$} and 2--10\,keV X-ray luminosities of the two nuclei are compared with the empirical AGN relation of \citet{2018ApJ...856..154S} in Figure~\ref{fig:shimizu}.
The relation of \citet{2018ApJ...856..154S} was originally calibrated using the absorption-corrected 14--150\,keV luminosities of \textit{Swift}/BAT-selected AGN. 
To facilitate comparison with our measurements, we converted the relation to the 2--10\,keV band assuming a power-law X-ray spectrum with photon index $\Gamma=1.8$, yielding a constant luminosity scaling between the two bands.
The broad \ion{H}{$\alpha$} luminosities have been corrected for BLR extinction using the broad-line Balmer decrements (Section~\ref{sec:BH_mass}). 
The corresponding observed luminosities are lower by approximately 0.7--0.9 dex.

Source~2 lies fully consistent with the local AGN relation within the intrinsic scatter, while Source~1 remains marginally offset toward higher X-ray luminosities at fixed broad \ion{H}{$\alpha$} luminosity. 
Nevertheless, both nuclei occupy the locus expected for local broad-line AGN, indicating broad agreement between their extinction-corrected broad-line and intrinsic X-ray luminosities. 
The modest offset of Source~1 is comparable to the intrinsic scatter of the \citet{2018ApJ...856..154S} relation and may reflect a combination of variability between the optical and X-ray observations, uncertainties in the extinction correction, or uncertainties in the intrinsic X-ray luminosity associated with the absorber geometry, column density, and assumed spectral model.

The weaker broad \ion{H}{$\beta$} emission relative to broad \ion{H}{$\alpha$} may indicate partial obscuration of the broad-line region, commonly observed in partially obscured Seyfert galaxies and intermediate-type AGN \citep[e.g.,][]{1993ApJ...414..552O,2008MNRAS.383..581D}. 
The broad-line Balmer decrements imply $E(B-V)_{\rm BLR}=0.687\pm0.121$ for Source~1 and $0.854\pm0.389$ for Source 2, confirming the presence of significant nuclear reddening. 
Consistent with the broad-line reddening measurements, the X-ray spectral properties also indicate significant nuclear obscuration in both nuclei, with $\log(N_{\rm H}/{\rm cm^{-2}})\simeq22.9$ and $\simeq23.6$ for Source~1 and Source~2, respectively.
Such column densities are consistent with the substantial obscuration frequently observed in merger-driven AGN, although they remain below the Compton-thick columns measured in some nearby dual-AGN systems \citep[e.g.,][]{2018Natur.563..214K,2023ApJ...954..116P}.
Comparing the X-ray column densities with the reddening inferred from the broad-line Balmer decrements yields $N_{\rm H}/E(B-V)$ ratios approximately 17 and 75 times larger than the canonical Galactic value \citep{1978ApJ...224..132B,2009MNRAS.400.2050G}.
Such elevated gas-to-dust ratios are commonly observed in Seyfert galaxies and other AGN \citep{2001A&A...365...28M,2001A&A...365...37M}, and are generally interpreted as evidence that a significant fraction of the absorbing gas resides in the circumnuclear AGN environment, potentially within or near the dust sublimation radius, rather than in the host-galaxy interstellar medium.

While recent studies suggest that Green Pea galaxies reside in low-density large-scale environments \citep[][]{2026A&A...710A.104G}, morphological studies increasingly support a connection between mergers and AGN triggering in compact star-forming galaxies.
For example, \citet{2026ApJS..282...14S} found that a substantial fraction of Green Pea candidates exhibit disturbed morphologies or merger signatures \citep[see also][]{2026A&A...711A..91A}.
J1622+3521, therefore, provides direct observational evidence that merger-driven SMBH growth can occur within the Green Pea population.

Recent studies have identified only a small number of candidate dual AGN systems in dwarf and low-mass galaxy mergers \citep[e.g.,][]{2023ApJ...944..160M,2024OJAp....7E...3M}. 
Compared with the broader dual-AGN population \citep[e.g.,][]{2009ApJ...698..956C,2017MNRAS.468.1273R,2018Natur.563..214K,2019MNRAS.487.2491E,2023MNRAS.519.5149D,2023ApJ...954..116P,2025ApJS..281...25P}, J1622+3521 is unusual not because it exhibits signatures of an ongoing interaction, but because of the nature of its host galaxy.
Like many confirmed dual AGN, the system shows dual nuclei and morphological signatures of an ongoing merger.
Its distinguishing feature is that both actively accreting SMBHs reside within a compact Green Pea galaxy characterized by intense star formation and extreme ionization conditions.
The detection of two actively accreting AGN in such an environment extends the known dual-AGN population into a host-galaxy regime that remains poorly represented in current samples.
Despite its comparatively modest stellar mass, both nuclei exhibit Seyfert-like X-ray luminosities and relatively high Eddington ratios, indicating efficient SMBH growth within a compact Green Pea environment.

Additionally, we estimated the expected contribution from high-mass X-ray binaries (HMXBs) using the empirical \(L_{\mathrm{X}}-\mathrm{SFR}-Z\) relation of \cite{2016MNRAS.457.4081B}. 
Adopting the total system SFR of \(1.12\,M_\odot\,\mathrm{yr}^{-1}\) and a representative metallicity for Green Pea galaxies of \(12+\log(\mathrm{O/H})=8.1\),\footnote{We do not adopt the metallicity reported in the MPA-JHU catalog because the integrated emission-line spectrum of J1622+3521 is dominated by AGN photoionisation. 
Consequently, metallicity estimates based on the integrated emission-line ratios may not reliably reflect the gas-phase abundance of the underlying star-forming component.} we obtain an expected \(0.5\)–\(8\) keV luminosity of \(L_{\mathrm{X,HMXB}}\sim7.7\times10^{39}\,\mathrm{erg\,s^{-1}}\). 
This is approximately four orders of magnitude below the intrinsic X-ray luminosities measured for the two sources. 
Even accounting for the intrinsic scatter of the empirical relation, the expected HMXB contribution remains negligible. 
We therefore conclude that X-ray binaries cannot account for the observed X-ray emission, either in observed or intrinsic luminosity, supporting an AGN origin for the detected X-ray sources.

\subsection{Black hole and host-galaxy properties}
\label{sec:BH_host_properties}

\begin{figure}
    \centering
    \includegraphics[width=\columnwidth]{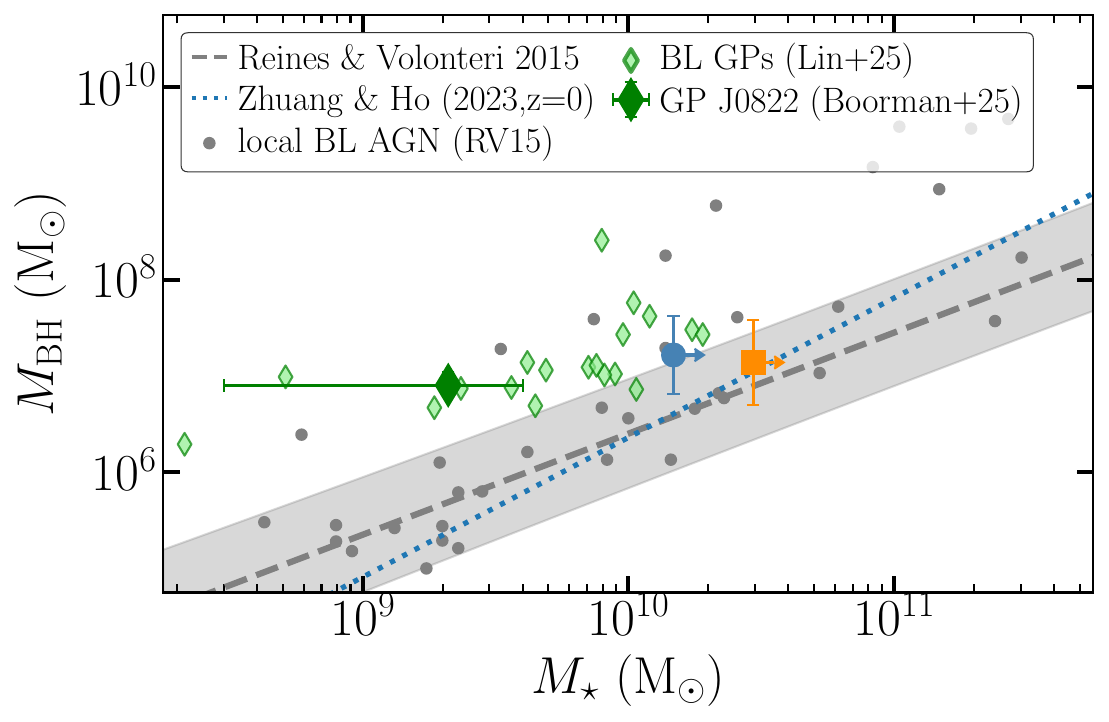}
    \caption{
    Black hole mass versus stellar mass for J1622+3521 compared with local supermassive black hole--host galaxy scaling relations. 
    The dashed gray line and shaded region indicate the best-fit relation and intrinsic scatter from \citet{2015ApJ...813...82R}, while the dotted blue line shows the AGN relation of \citet{2023NatAs...7.1376Z} evolved to $z=0$. 
    Gray circles represent the local broad-line AGN sample compiled by \citet{2015ApJ...813...82R}. 
    Light green diamonds show the broad-line Green Pea (BLGP) galaxies identified by \citet{2025ApJ...980L..34L}. 
    The blue and orange symbols correspond to Source~1 and Source~2 of J1622+3521.
    The green point marks the comparison AGN Green Pea J0822+2241 from \citet{2025ApJ...988..157B}. 
    }
    \label{fig:Reines_Volenteri}
\end{figure}

The inferred supermassive black hole and stellar masses are shown in Figure~\ref{fig:Reines_Volenteri}, where we compare J1622+3521 with the local supermassive black hole--stellar mass relations of \citet{2015ApJ...813...82R} and \citet{2023NatAs...7.1376Z}.
The stellar masses were obtained from the FADO spectral decomposition of the DEIMOS spectra, yielding $M_{\star,1} = 1.48\times10^{10}\,M_\odot$ and $M_{\star,2} = 2.95\times10^{10}\,M_\odot$ for Sources~1 and 2, respectively.
These estimates correspond to the stellar populations sampled by the spectroscopic extraction apertures associated with each nucleus, rather than to integrated stellar masses of the interacting galaxies.
Consequently, they may underestimate the total stellar masses associated with the two galaxies.
The compact morphology of the system suggests that this aperture effect may be modest, although its magnitude is not directly constrained by the present measurements.
The inferred black hole masses are less sensitive to aperture effects, as they are derived from the nuclear broad-line emission.
Therefore, the true locations of the two nuclei in the $M_{\rm BH}$--$M_\star$ plane could shift toward higher stellar masses, moving them to the right in Figure~\ref{fig:Reines_Volenteri}.

Within the uncertainties, both nuclei lie broadly consistent with the loci defined by nearby AGN and galaxies despite the compact, intensely star-forming, and interacting nature of the Green Pea system.
The inferred black hole masses place both sources close to the local scaling relations of \citet{2015ApJ...813...82R} and \citet{2023NatAs...7.1376Z}.
Although many of the broad-line Green Pea galaxies identified by \citet{2025ApJ...980L..34L} appear offset toward higher black hole masses at fixed stellar mass, the two nuclei of J1622+3521 remain consistent with the overall scatter of the broad-line Green Pea population while residing closer to the local scaling relations.
The comparison AGN Green Pea J0822+2241 \citep{2025ApJ...988..157B} lies within the scatter of the broad-line Green Pea population and is similarly offset above the local scaling relations.
Taken together, these comparisons suggest that compact Green Pea galaxies can host rapidly accreting supermassive black holes spanning a range of locations relative to the local black hole--host galaxy scaling relations. 
The proximity of J1622+3521 to the local relations may indicate that black hole growth and stellar mass assembly have remained broadly coupled in this system despite the ongoing interaction, although systematic uncertainties in both virial black hole masses and stellar-mass estimates should be kept in mind.

\subsection{Comparison with Compact High-Redshift AGN}
\label{sec:Comp_with_highz_AGN}

The properties of J1622+3521 can be considered in the broader context of compact AGN populations identified at high redshift.
Both nuclei exhibit broad \ion{H}{$\alpha$} emission with FWHM $\sim1650$--$1900$~km~s$^{-1}$, compact morphology, and significant nuclear obscuration inferred from the X-ray spectral fitting.
The measured column densities, $\log(N_{\rm H}/{\rm cm^{-2}})\simeq22.9$ for Source~1 and $\simeq23.6$ for Source~2, indicate substantial obscuration toward both nuclei.
These properties provide a useful comparison with compact broad-line AGN identified at high redshift, some of which also exhibit broad Balmer emission and substantial nuclear obscuration \citep[e.g.,][]{2023ApJ...954L...4K}.

The UV-to-optical continuum of J1622+3521 provides a complementary point of comparison with these populations.
The unresolved SED exhibits a V-shaped continuum, with a blue UV component and a positive but relatively shallow optical slope (Section~\ref{sec:optical_uv_continuum}).
The spectroscopic UV slope, $\beta_{\rm UV}=-1.1$, falls within the range reported for LBDs, compact broad-line AGN identified with \textit{JWST} \citep{2026arXiv260322277S}.
The positive optical slope, $\beta_{\rm opt}=0.25$, is consistent with the redward-rising optical continuum commonly associated with the V-shaped SEDs of LRDs \citep{2025ApJ...986..126K}.
Although LBDs and LRDs share several structural and spectroscopic characteristics, LBDs exhibit substantially bluer continua \citep{2026arXiv260322277S}.
Thus, the UV slope of J1622+3521 overlaps the reported LBD range, while its integrated UV-to-optical continuum is more closely characterized by an LRD-like V-shaped SED.
However, the continuum measurements include the combined contribution of both nuclei and cannot establish whether either nucleus independently satisfies the corresponding continuum criteria.
We therefore describe J1622+3521 as a local system with phenomenological similarities to these compact AGN populations, rather than classifying it as an LRD or LBD analog.

The comparison with local Green Pea AGN is also informative.
Green Pea galaxies are compact, intensely star-forming systems that provide nearby laboratories for studying some of the physical conditions encountered in high-redshift star-forming galaxies.
Recent X-ray studies have revealed evidence for obscured AGN activity in at least some Green Pea galaxies, including luminous hard X-ray emission and moderate nuclear obscuration \citep[e.g.,][]{2025ApJ...988..157B}.
Preliminary work (Maillard et al., in preparation) further suggests that some local compact intensely star-forming galaxies \citep{2011ApJ...728..161I} may occupy a photometric parameter space similar to that of LBDs.
J1622+3521 adds to this picture by demonstrating that dual AGN activity can occur in a compact, intensely star-forming Green Pea system.

The stronger obscuration of Source~2 is particularly noteworthy.
Its larger X-ray column density and broad-line Balmer decrement, corresponding to $E(B-V)_{\rm BLR}=0.86$, indicate substantial attenuation of the BLR emission.
This provides a point of comparison with the reddening and Balmer-line properties reported for some compact high-redshift AGN and LRD candidates \citep[e.g.,][]{2024ApJ...963..129M,2024ApJ...976...96P}.
The relative weakness of broad \ion{H}{$\beta$} compared to broad \ion{H}{$\alpha$} is likewise consistent with partially obscured BLR emission.
Similar Balmer-line properties are observed in Type~1.9 AGN \citep[e.g.,][]{2018ApJ...856..154S} and have also been reported in compact broad-line AGN identified with \textit{JWST} at high redshift \citep[e.g.,][]{2025ApJ...986..177B}.
At the same time, the spectral modeling does not require complete suppression of the nuclear X-ray emission despite the substantial obscuration.
In this respect, the coexistence of broad Balmer emission and detectable X-rays in J1622+3521 provides a phenomenological comparison with the X-ray Dot \citep{2026ApJ..1000L..18H}, a compact high-redshift source that exhibits several LRD-like properties while also hosting luminous X-ray emission.

We additionally compare J1622+3521 with compact AGN populations identified at high redshift using the broad \ion{H}{$\alpha$}--X-ray plane shown in Figure~\ref{fig:shimizu}.
Several compact AGN candidates identified with \textit{JWST} have been reported to be X-ray weak relative to their optical or bolometric luminosities \citep[e.g.,][]{2025MNRAS.538.1921M,2026A&A...706A.302C}, whereas both nuclei in J1622+3521 remain broadly consistent with the local broad-line AGN relation of \citet{2018ApJ...856..154S}.
In this parameter space, the two nuclei are more consistent with the X-ray-normal regime represented by the X-ray Dot than with the X-ray-weak compact AGN candidates shown in Figure~\ref{fig:shimizu}.

A complementary comparison is provided by the recent \textit{Einstein Probe} study of six V-shaped broad-line Green Pea galaxies selected as local analogs of LRDs by \citet{2026arXiv261032430C}.
Their sample exhibits a range of X-ray properties, with two sources consistent with or above standard Type~1 AGN relations, while the stacked X-ray-weak subsample lies approximately 0.5--0.7~dex below the expected X-ray luminosities at fixed broad-line or UV luminosity.
The two nuclei of J1622+3521 are consistent with the X-ray-normal part of this comparison sample, with Source~2 close to the local AGN relation and Source~1 showing a modest offset toward higher X-ray luminosity.
Together, these comparisons illustrate the diversity of X-ray properties among compact broad-line AGN and Green Pea systems, ranging from moderately X-ray-weak sources to objects broadly consistent with standard AGN expectations.

A complementary comparison can also be made using the black hole--host galaxy mass properties of broad-line LRDs identified at high redshift.
\citet{2025ApJ...986..126K} found that their broad-line LRD sample generally lies above the local $M_{\rm BH}$--$M_\star$ relation, with elevated black-hole-to-host mass ratios compared with nearby broad-line AGN.
In contrast, the two nuclei of J1622+3521 appear closer to the local scaling relations in the adopted mass estimates (Section~\ref{sec:BH_host_properties}).
This suggests that their black-hole-to-host mass ratios may be less extreme than those reported for many broad-line LRDs, although the comparison is subject to systematic differences in the methods used to estimate the stellar masses.
In particular, the stellar masses of J1622+3521 are derived from FADO fits to the DEIMOS extraction spectra and may not represent the integrated stellar masses of the interacting galaxies.

The agreement between the observed \ion{He}{II} and soft X-ray luminosities and the empirical AGN relation of \citet{2025A&A...696A.133K} provides a further, complementary consistency check.
Both nuclei lie close to this relation and remain consistent with its intrinsic scatter, despite their optical reddening and substantial nuclear obscuration.
This agreement supports the interpretation that the observed narrow \ion{He}{II} emission is associated with the ionizing radiation from the AGN in both nuclei.
While the spectral modeling favors moderate-to-high obscuration, the present data cannot completely exclude more heavily obscured geometries in which a fraction of the observed X-ray emission arises from scattered or reflected light.
Future hard X-ray observations will be important for constraining the presence of any deeply buried AGN components.

Overall, J1622+3521 provides a local example of dual AGN activity in a compact, intensely star-forming environment.
Its broad-line emission, nuclear obscuration, and unresolved UV-to-optical continuum offer complementary points of comparison with compact AGN populations identified at high redshift.
At the same time, J1622+3521 differs from many of these systems in several respects: both nuclei are detected in X-rays and remain broadly consistent with local AGN relations, while their inferred black-hole-to-host mass ratios appear closer to local scaling relations than those reported for many broad-line LRDs.
The system's dual nature and Green Pea host environment further distinguish it from the individual compact AGN commonly identified at high redshift.
Although the present data do not establish an LBD- or LRD-like continuum for either nucleus individually, J1622+3521 provides a useful local reference for investigating both the similarities and the diversity of compact, obscured, and rapidly accreting AGN across cosmic time.

\section{Conclusions}
\label{sec:conclusion}

We report the discovery and multiwavelength characterization of a dual AGN hosted within the Green Pea system J1622+3521. 
The system contains two spatially distinct X-ray and optical nuclei separated by $\sim8.4$~kpc, both of which exhibit clear signatures of active supermassive black hole accretion. 
The combined X-ray and optical spectroscopic analysis establishes J1622+3521 as a rare and clearly resolved example of a dual AGN hosted within a Green Pea-like environment.

The main results of this work can be summarized as follows:

\begin{enumerate}

\item Both nuclei are luminous X-ray sources with intrinsic luminosities of $\log L_{2-10\,{\rm keV}}\simeq43.9$ and $43.6$~erg~s$^{-1}$ for Source~1 and Source~2, respectively, confirming the presence of two actively accreting supermassive black holes. 
The X-ray spectra indicate significant obscuration, with column densities of $\log(N_{\rm H}/{\rm cm^{-2}})\sim22.9$ and $\sim23.6$, while remaining below the Compton-thick regime commonly observed in heavily obscured merger systems.

\item Optical spectral analysis reveals broad \ion{H}{$\alpha$} emission in both nuclei, with FWHM $\sim1700$--$1900$~km~s$^{-1}$, as well as high-ionization emission lines, including \ion{He}{II} and [\ion{Fe}{VII}] in the DEIMOS spectra and [\ion{Ne}{V}] in the SDSS fibre spectrum.
The narrow-line ratios place both nuclei within the AGN region of the BPT diagram, supporting the dual-AGN interpretation.

\item Virial supermassive black hole mass estimates based on the broad \ion{H}{$\alpha$} emission imply black hole masses of $\log(M_{\rm BH}/M_\odot)\sim7.3$ for both Source~1 and Source~2.
Combined with the intrinsic X-ray luminosities, these estimates imply efficient accretion, with Source~1 accreting at $\lambda_{\rm Edd}\sim0.7$ and Source~2 at $\lambda_{\rm Edd}\sim0.3$.

\item The combination of compact morphology, intense nebular emission, significant obscuration, and efficient supermassive black hole accretion is consistent with the possibility that the ongoing interaction is facilitating gas inflows toward the nuclear regions. 
Compared with the broader dual AGN population, J1622+3521 occupies a relatively unusual parameter space associated with compact, intensely star-forming galaxies rather than the massive infrared-luminous mergers that dominate many nearby dual AGN samples.

\item The unresolved UV-to-optical continuum exhibits a V-shaped spectral energy distribution, with a UV slope within the range commonly reported for LRDs and a positive but relatively shallow optical slope.
Together with the compact morphology, broad-line emission in both nuclei, and efficient accretion, these properties highlight phenomenological similarities between J1622+3521 and compact broad-line AGN populations identified at high redshift.
While the available data do not allow us to establish whether either nucleus independently satisfies the LRD continuum criteria, J1622+3521 provides an interesting nearby reference for studying the continuum properties of compact, rapidly accreting AGN.
The detectable X-ray emission from both nuclei also distinguishes this system from many X-ray-weak or X-ray-undetected compact AGN candidates reported in recent \textit{JWST} studies.

\end{enumerate}

Overall, J1622+3521 demonstrates that efficient supermassive black hole growth and dual AGN activity can occur within compact, low-metallicity, and intensely star-forming galaxies. 
The system provides a nearby laboratory for investigating the connection between mergers, star formation, and black hole growth in an interacting, compact Green Pea-like environment. 
Future spatially resolved and multiwavelength observations, including integral field unit spectroscopy and observations with facilities such as \textit{JWST}, \textit{NewAthena}, ALMA, and VLBI interferometry, will be crucial for disentangling the contributions from AGN activity, star formation, and outflows, resolving the nuclear structure, constraining the obscuring gas geometry, and probing the connection between mergers, star formation, and supermassive black hole growth across cosmic time.

\begin{table*}
\centering
\caption{
Emission-line fluxes, full widths at half maximum (FWHMs), and luminosities for the two spectra of J1622+3521.
For [O\,III], the FWHMs are measured from the single-Gaussian (M1) fits, while the fluxes and luminosities are summed over the three narrow Gaussian components of the M3 fits.
Luminosities are calculated assuming a luminosity distance of $D_L=1400$ Mpc.
}
\label{tab:balmer_results_compact}
\begin{tabular}{l|ccc|ccc}
 & \multicolumn{3}{c}{S1} & \multicolumn{3}{c}{S2} \\
Emission line
& Flux
& FWHM
& Luminosity
& Flux
& FWHM
& Luminosity\\
& ($\rm 10^{-15} erg ~ s^{-1} ~ cm^{-2}$) & (km s$^{-1}$) & ($\rm 10^{41} erg ~ s^{-1}$) & ($\rm 10^{-15} erg ~ s^{-1} ~ cm^{-2}$) & (km s$^{-1}$) & ($\rm 10^{41} erg ~ s^{-1}$) \\
\hline
He\,II $\lambda4686$
& $0.255 \pm 0.009$ & $309^{+4}_{-9}$ & $0.598 \pm 0.020$
& $0.078 \pm 0.005$ & $309^{+11}_{-12}$ & $0.183 \pm 0.011$ \\

H$\beta$ (narrow)
& $1.392 \pm 0.027$ & $328^{+3}_{-5}$ & $3.265 \pm 0.062$
& $0.310 \pm 0.041$ & $322^{+7}_{-66}$ & $0.726 \pm 0.097$ \\

H$\beta$ (broad)
& $0.294 \pm 0.038$ & $939^{+33}_{-62}$ & $0.690 \pm 0.088$
& $0.204 \pm 0.093$ & $756^{+43}_{-43}$ & $0.480 \pm 0.218$ \\

[O\,III] $\lambda4959$
& $3.227 \pm 0.485$ & $411^{+4}_{-8}$ & $7.567 \pm 1.137$
& $0.723 \pm 0.056$ & $549^{+7}_{-12}$ & $1.695 \pm 0.131$ \\

[O\,III] $\lambda5007$
& $9.680 \pm 1.454$ & $411^{+4}_{-8}$ & $22.701 \pm 3.410$
& $2.168 \pm 0.167$ & $549^{+7}_{-12}$ & $5.085 \pm 0.392$ \\

[N\,II] $\lambda6548$
& $0.987 \pm 0.033$ & $330^{+17}_{-3}$ & $2.265 \pm 0.077$
& $0.589 \pm 0.043$ & $579^{+10}_{-73}$ & $1.364 \pm 0.100$ \\

H$\alpha$ (narrow)
& $4.816 \pm 0.168$ & $330^{+17}_{-3}$ & $11.295 \pm 0.393$
& $2.911 \pm 0.213$ & $579^{+10}_{-73}$ & $6.826 \pm 0.500$ \\

H$\alpha$ (broad)
& $2.037 \pm 0.124$ & $1894^{+56}_{-56}$ & $4.777 \pm 0.290$
& $1.721 \pm 0.047$ & $1655^{+36}_{-36}$ & $4.035 \pm 0.111$ \\

[N\,II] $\lambda6583$
& $2.897 \pm 0.098$ & $330^{+17}_{-3}$ & $6.795 \pm 0.231$
& $1.745 \pm 0.128$ & $579^{+10}_{-73}$ & $4.091 \pm 0.299$ \\

[Fe VII] $\lambda$6087
    & $0.0339^{+0.0048}_{-0.0004}$
    & $200^{+3}_{-17}$
    & $0.0794^{+0.0113}_{-0.0010}$
    & $0.0117^{+0.0033}_{-0.0006}$
    & $159^{+2}_{-87}$
    & $0.0274^{+0.0078}_{-0.0014}$ \\

\hline
\multicolumn{7}{l}{\textit{SDSS spectrum (combined S1+S2)}} \\

\multicolumn{1}{l|}{[Ne\,V] $\lambda3346$}
& \multicolumn{6}{c}{$0.188 \pm 0.029$ (Flux), 
$126^{+18}_{-17}$ (FWHM), 
$0.440 \pm 0.067$ (Luminosity)} \\

\multicolumn{1}{l|}{[Ne\,V] $\lambda3426$}
& \multicolumn{6}{c}{$0.496 \pm 0.051$ (Flux), 
$123^{+18}_{-17}$ (FWHM), 
$1.163 \pm 0.120$ (Luminosity)} \\

\end{tabular}
\end{table*}

\begin{acknowledgments}

K.K. acknowledges support from the institutional project RVO:67985815 and from the INTER-COST LUC24023 project of the INTER-EXCELLENCE II programme funded by the Czech Ministry of Education, Youth and Sports.
J.S. acknowledges discussions with Juan Antonio Fernandéz Ontiveros and Teo~Mu\~noz-Darias during his visit at the IAC. M.G. and J.S. ackonowledge the financial support from the Czech Science Foundation (GAČR) project 26-22614S.
The work of DS was carried out at the Jet Propulsion Laboratory, California Institute of Technology, under a contract with the National Aeronautics and Space Administration (80NM0018D0004).
This research has made use of the Keck Observatory Archive (KOA), which is operated by the W. M. Keck Observatory and the NASA Exoplanet Science Institute under contract with the National Aeronautics and Space Administration.
This research has made use of the NASA/IPAC Extragalactic Database (NED), which is funded by the National Aeronautics and Space Administration and operated by the California Institute of Technology.
This research has made use of NASA's Astrophysics Data System Bibliographic Services.
This research has made use of the SIMBAD database, operated at CDS, Strasbourg, France.
This work made extensive use of the Python packages NumPy, Matplotlib, SciPy, and Astropy.

\end{acknowledgments}


\section*{Data availability}

The \textit{Chandra} observations analysed in this study are publicly available through the Chandra Data Archive under Observation IDs 28142, 30637, and 30638 and are contained in Chandra Data Collection (CDC) 648~\href{https://doi.org/10.25574/cdc.648}{doi:10.25574/cdc.648}.
The Sloan Digital Sky Survey (SDSS) and DESI Legacy Imaging Surveys (DECaLS) data used in this work are publicly available through their respective archives.
The Keck/DEIMOS observations are available through the Keck Observatory Archive (KOA).
The GALEX GR6/GR7 catalogue data used for the UV photometry are available from MAST through the GALEX MCAT catalogue: \href{https://doi.org/10.17909/T9H59D}{doi:10.17909/T9H59D}.
The AllWISE Source Catalog photometry used in this work is publicly available through the NASA/IPAC Infrared Science Archive (IRSA) under DOI \href{https://doi.org/10.26131/IRSA1}{10.26131/IRSA1}.
Derived data products generated during this study are available from the corresponding authors upon reasonable request.

\section*{Code availability}

This study made use of publicly available software, including CIAO, Sherpa, PypeIt, FADO, BXA, XSPEC/PyXspec, Astropy, SciPy, NumPy, and Matplotlib.
Custom scripts used for data reduction, analysis, and figure production are available from the corresponding authors upon reasonable request.

\appendix

\section{High-ionization emission lines}
\label{app:high_ionization}

\begin{figure*}
    \centering

    \makebox[\textwidth][c]{%
    \includegraphics[width=0.32\textwidth]{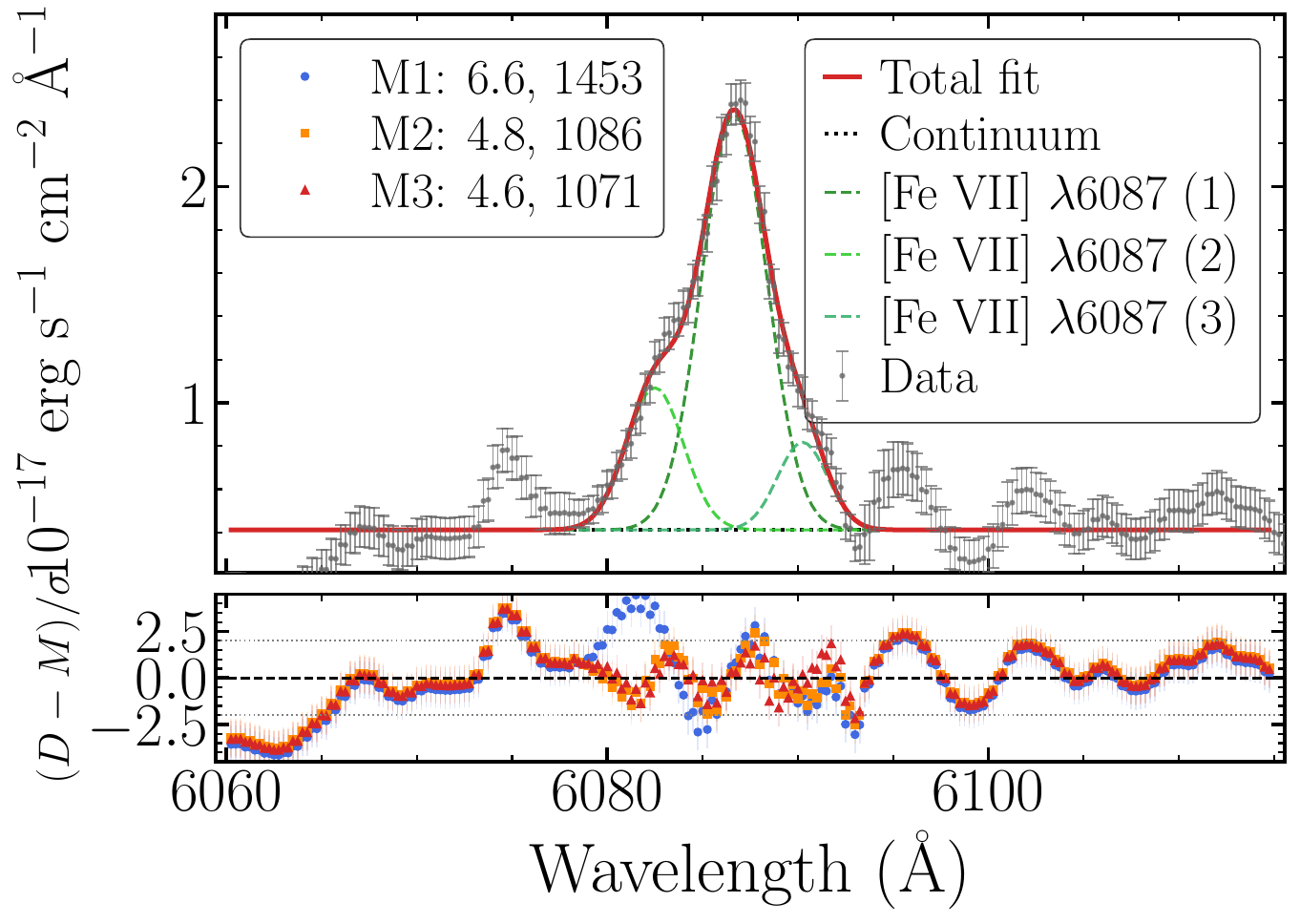}
    \includegraphics[width=0.32\textwidth]{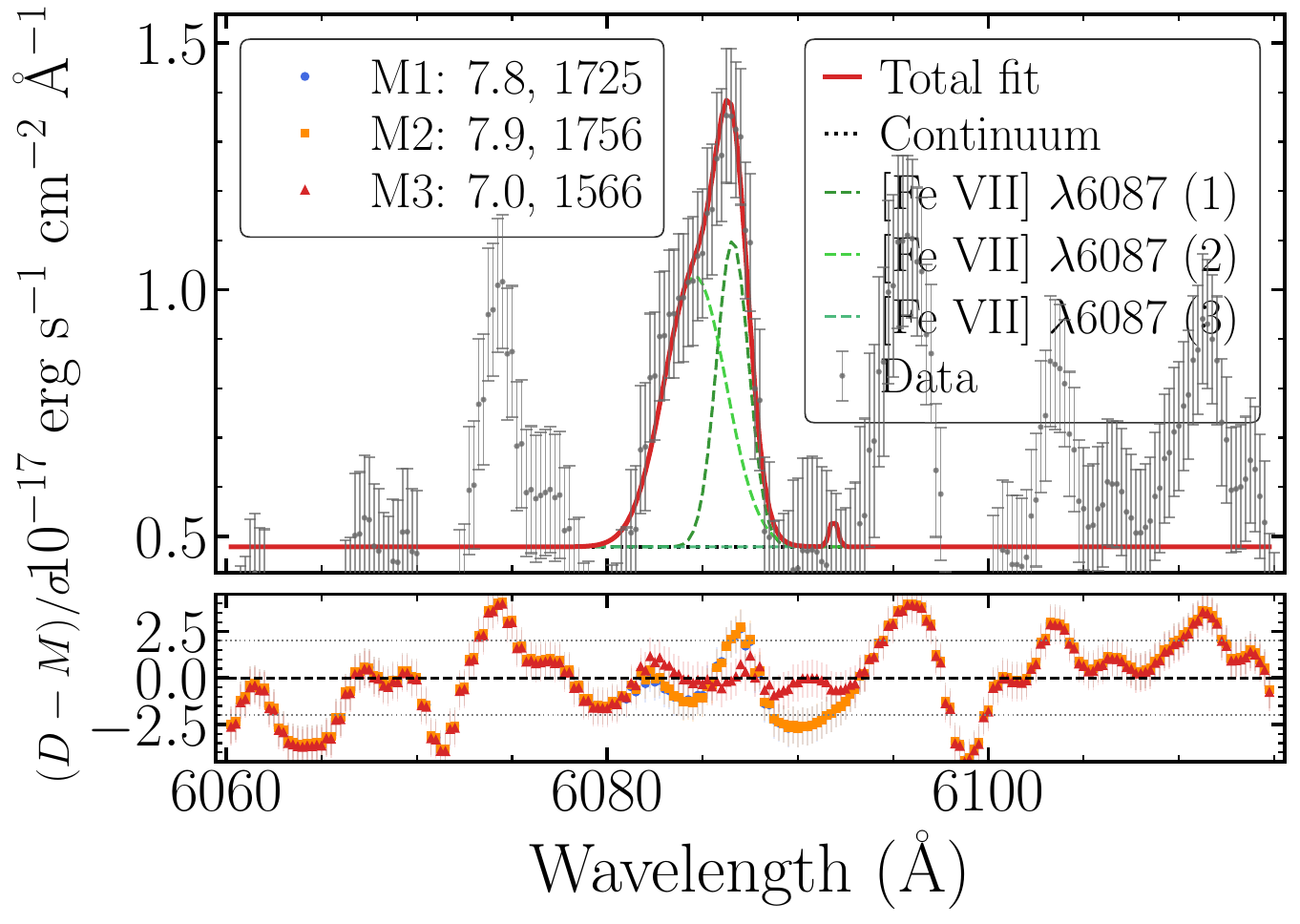}
    \includegraphics[width=0.32\textwidth]{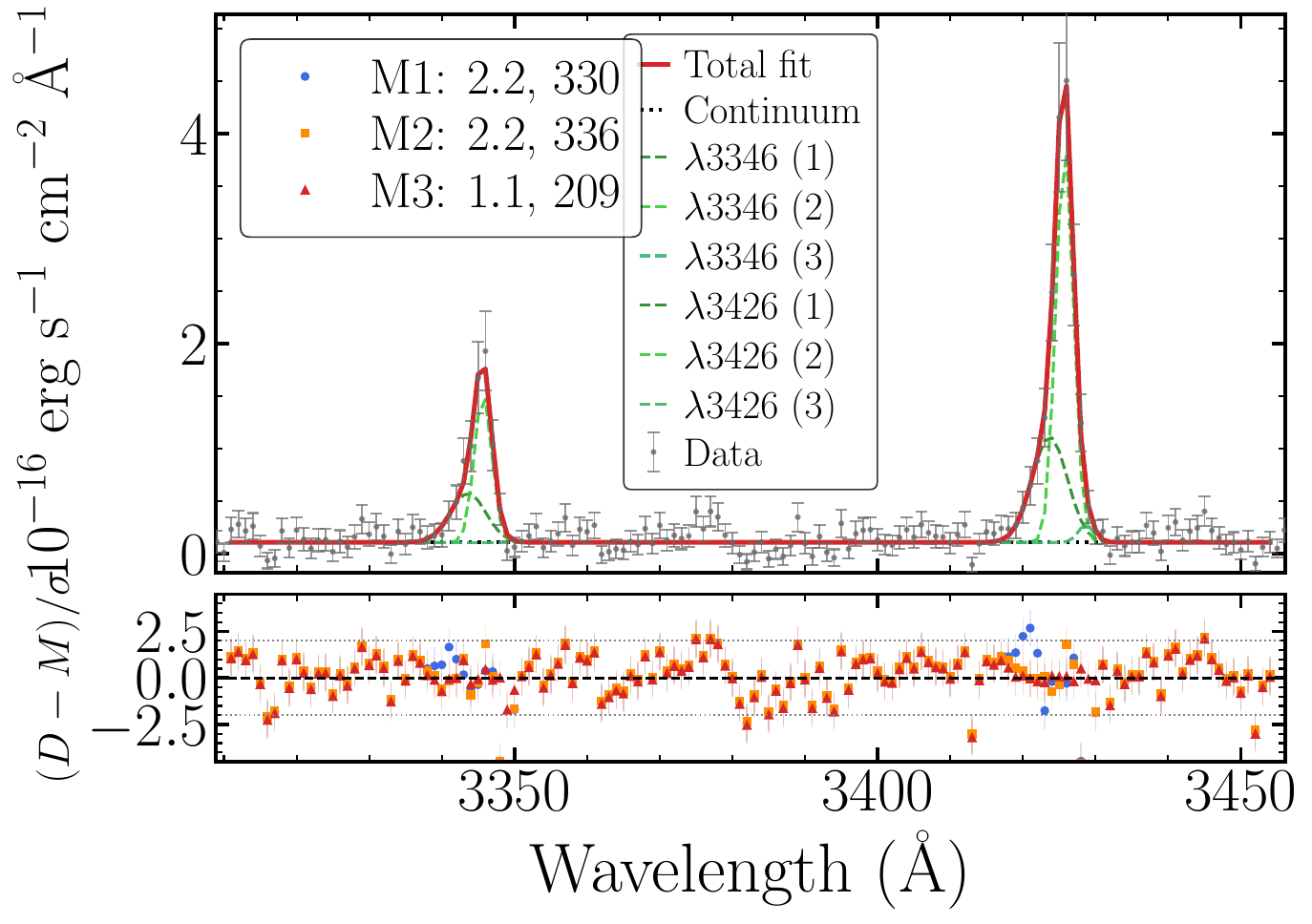}
    }

    \caption{
        Multi-component Gaussian fits to the high-ionization [\ion{Fe}{VII}] and [\ion{Ne}{V}] emission lines in J1622+3521.
        From left to right, the panels show the [\ion{Fe}{VII}] $\lambda6087$ emission in the Keck/DEIMOS spectra of Source~1 and Source~2, and the [\ion{Ne}{V}] $\lambda\lambda3346,3426$ doublet in the SDSS fibre spectrum.
        The red curve represents the total best-fit M3 model, the dotted black line marks the fitted continuum, and the dashed curves show the individual Gaussian components.
        The [\ion{Ne}{V}] doublet is fitted simultaneously, with independent amplitudes for the two transitions.
        The lower panel in each spectrum shows the normalized residuals, $(\mathrm{Data}-\mathrm{Model})/\sigma$, for the M1, M2, and M3 models, overlaid using different colors and marker symbols.
        The reduced $\chi^2$ and approximate Bayesian information criterion (BIC) are reported in the legend for each model.
        These fits are presented to illustrate the detection of high-ionization emission lines and are not used in the subsequent quantitative analysis.
    }
    \label{fig:coronal_fittings}
\end{figure*}

In addition to the emission lines discussed in the main text, we detect high-ionization [\ion{Fe}{VII}] $\lambda6087$ emission in both Keck/DEIMOS spectra of J1622+3521, as well as the [\ion{Ne}{V}] $\lambda\lambda3346,3426$ doublet in the SDSS fibre spectrum.
The [\ion{Ne}{V}] doublet is not covered by the wavelength range of the Keck/DEIMOS spectra and is therefore only available in the SDSS spectrum.
Given the size and placement of the SDSS fibre, which encompasses both nuclei, we attribute this emission to the combined contribution of Source~1 and Source~2.
These transitions require highly energetic ionizing photons and provide further evidence for the presence of active galactic nuclei in the system.
We present the corresponding spectral fits in Figure~\ref{fig:coronal_fittings}.

The line profiles are fitted using multiple Gaussian components to assess their detection and account for possible profile complexity.
The [\ion{Fe}{VII}] $\lambda6087$ line is fitted independently in the spectra of Source~1 and Source~2, while the two [\ion{Ne}{V}] transitions are fitted simultaneously, with their amplitudes allowed to vary independently.
Although the [\ion{Fe}{VII}] $\lambda6087$ feature is fitted in both spectra, we do not consider it to be firmly detected in Source~2.
In particular, the multi-component fits do not provide sufficient statistical and spectroscopic evidence for a reliable detection, despite the improvement in BIC relative to simpler models.
The resulting fit also retains substantial residual structure, indicating that the line profile is not adequately described by the adopted model.
We therefore regard the [\ion{Fe}{VII}] emission in Source~2 as tentative.
These measurements are included to document the presence of high-ionization emission in J1622+3521, but are not used in the quantitative analysis presented elsewhere in this work.

\bibliographystyle{aasjournalv7}
\bibliography{bibliography}

\end{document}